\documentclass[journal,comsoc]{IEEEtran}

\usepackage[T1]{fontenc}% optional T1 font encoding
\usepackage{graphicx}
\usepackage{cite}
\usepackage{picinpar}
\usepackage{amsmath}
\usepackage{amssymb,bm}
\usepackage{float}
\usepackage{stfloats}
\usepackage{url}
\usepackage[latin1]{inputenc}
\usepackage{colortbl}
\usepackage{soul}
\usepackage{multirow}
\usepackage{pifont}
\usepackage{color}
\usepackage{alltt}
\usepackage{enumerate}
\usepackage{siunitx}
\usepackage{hyperref} 
\usepackage{breakurl}
\usepackage{epstopdf}
\usepackage{pbox}
\usepackage[linesnumbered,ruled]{algorithm2e}
\usepackage{comment}
\usepackage{amsthm}
\theoremstyle{plain}
\newtheorem{thm}{Theorem}
\newtheorem{lem}{Lemma}
\newtheorem{rem}{Remark}
\usepackage{subfig}
\usepackage{lipsum}% For this example
\usepackage{titlesec}
\usepackage{setspace}

\begin{document}
%
% The following two lines reduce the space between text and equations
 \addtolength\abovedisplayskip{-0.35\baselineskip}%
 \addtolength\belowdisplayskip{-0.35\baselineskip}%
 %The following three lines reduce the space between text and headings of section and subsections
\titlespacing{\section}{0em}{.6em}{.5em}
\titlespacing{\subsection}{0em}{.5em}{.25em}
\titlespacing{\subsubsection}{0em}{.5em}{.25em}
 % The following set space after figure and before text
\setlength{\textfloatsep}{3pt}

\title{Design and Analysis of Clustering-based Joint Channel Estimation and Signal Detection for NOMA}

\author{\IEEEauthorblockN{Ayoob~Salari,~\IEEEmembership{Graduate~Student~Member,~IEEE},
		Mahyar~Shirvanimoghaddam,~\IEEEmembership{Senior~Member,~IEEE}, Muhammad~Basit~Shahab,~\IEEEmembership{Member,~IEEE}, Reza Arablouei, Sarah~Johnson,~\IEEEmembership{Senior~Member,~IEEE}
		\vspace{-1.9em}}
\thanks{Ayoob Salari and Mahyar Shirvanimoghaddam are with the School of Electrical and Information Engineering, The University of Sydney, Camperdown, NSW 2006, Australia (e-mail: ayoob.salari@sydney.edu.au; mahyar.shirvanimoghaddam@sydney.edu.au)}
\thanks{Muhammad Basit Shahab and Sarah Johnson are with the School of Engineering, University of Newcastle, Callaghan, NSW 2308, Australia (e-mail: basit.shahab@newcastle.edu.au; sarah.johnson@newcastle.edu.au).}
\thanks{Reza Arablouei is with the Commonwealth Scientific and Industrial Research Organisation, Pullenvale, QLD 4069, Australia (e-mail:
reza.arablouei@csiro.au).}
}

\maketitle

\begin{abstract}
We propose a joint channel estimation and signal detection approach for the uplink non-orthogonal multiple access (NOMA) using unsupervised machine learning. We apply a Gaussian mixture model (GMM) to cluster the received signals, and accordingly optimize the decision regions to enhance the symbol error rate (SER) performance. We show that, when the received powers of the users are sufficiently different, the proposed clustering-based approach achieves an SER performance on a par with that of the conventional maximum-likelihood detector (MLD) with full channel state information (CSI). We study the tradeoff between the accuracy of the proposed approach and the blocklength, as the accuracy of the utilized clustering algorithm depends on the number of symbols available at the receiver. We provide a comprehensive performance analysis of the proposed approach and derive a theoretical bound on its SER performance. Our simulation results corroborate the effectiveness of the proposed approach and verify that the calculated theoretical bound can predict the SER performance of the proposed approach well. We further explore the application of the proposed approach to a practical grant-free NOMA scenario, and show that its performance is very close to that of the optimal MLD with full CSI, which usually requires long pilot sequences.
\end{abstract}

% Note that keywords are not normally used for peerreview papers.
\IEEEpeerreviewmaketitle
\vspace{-1em}
\section{Introduction}\label{introduction}
\IEEEPARstart{T}{he} fifth generation (5G) of mobile standards has introduced two main service categories, namely, massive machine-type communications (mMTC) and ultra-reliable low-latency communications (URLLC)~\cite{series2017minimum}, alongside the enhanced mobile broadband (eMBB) that has been the primary focus of all previous generations. The main emphasis of mMTC is on the handling of a huge number of low-throughput, delay-tolerant, energy-efficient, and low-cost devices, which are usually used in Internet of things (IoT) applications. A recent forecast estimates that the number of enterprise and automotive IoT devices will grow from 5.8 billion in 2020 to 41.6 billion in 2025 generating 79.4 zettabytes of data per annum~\cite{Framingham2019Growth}. While 5G is expected to serve many IoT applications, major breakthroughs in designing communication protocols and radio resource management techniques are required to serve applications with a diverse range of requirements in terms of data rate, reliability, availability, end-to-end latency, energy efficiency, security, and privacy~\cite{6GCellular}.

%To efficiently manage high density devices in massive IoT scenario while keeping the QoS criteria above the target, network should optimize the resource utilization~\cite{mahmood2020white}. \par
\subsection{Related Work}

Non-orthogonal multiple access (NOMA) schemes have gained significant attention in the last decade as an enabler for mMTC \cite{shahab2019grantfree}. NOMA, either power-domain or code-domain \cite{aldababsa2018tutorial},  allows multiple users to share the same radio resources,  leading to higher spectral efficiency. While in power-domain NOMA users transmit with different power levels depending on their channel conditions, code-domain NOMA relies on assigning a unique code to each user. 
%Some prominent code domain NOMA techniques are sparse code multiple access (SCMA)~\cite{ma2019sparse}, multi-user shared access (MUSA)~\cite{catak2019multi}, interleave-division multiple access (IDMA)~\cite{wu2018comprehensive}, and low-density spreading (LDS)~\cite{vaezi2019multiple}. There are also other NOMA techniques such as bit division multiplexing (BDM)~\cite{duan2020optimized} and pattern division multiple access (PDMA)~\cite{dai2018pattern}. Power-domain NOMA is compatible with the current communication networks and can improve spectral efficiency over the same bandwidth. Code-domain NOMA requires higher bandwidth and significant changes in the current systems \cite{dai2015non}.
%
%
Not only does NOMA allow multiple devices to share the same radio resources, but it can also reduce the signaling overhead and latency by allowing grant-free uplink connection \cite{abbas2020grantfree}.
%Current wireless networks allocate data transmission slots to users through a process called random access that is a multi-step handshake between the base station (BS) and the user. When there are many users, which is the case in massive IoT, this grant-based access suffers from excessive signaling overhead that take up significant resources to establish a connection. This problem is even more challenging in the grant-free access since we also need to consider access collisions and we do not have sufficient knowledge regarding the participant users. Subsequently, it can result in a substantial efficiency loss since the typical data size is comparable to the overhead signal. 
Incorporation of NOMA with the grant-free access that is a lightweight random access protocol is considered to be a key enabler of massive connectivity in IoT \cite{shahab2019grantfree}. 

In the massive IoT served by 5G and beyond, the communication network will be highly complex and dynamic with numerous possibly contending design requirements \cite{mahmood2020six,letaief2019roadmap}. Machine learning can be effectively used to enhance network intelligence and enable self-adaptability in an efficient and timely manner \cite{rodrigues2019edge}. Machine learning can be used to render effective and efficient services in different network layers. In the physical layer, clustering methods, such as $k$-means and Gaussian mixture model (GMM) can be utilized for channel estimation and signal detection while convolutional neural network algorithms can be applied for channel decoding \cite{kim2021deep}. To increase reliability, powerful supervised machine learning algorithms such as deep neural networks can be employed at the data link layer for user scheduling. Moreover, reinforcement learning can be utilized at the network layer to improve network robustness and service continuity \cite{tang2019future,Muhammad2020Artificial}. 

Channel estimation schemes can be mainly categorized into training-based \cite{zhang2006optimal}, blind \cite{shin2007blind}, and semi-blind \cite{liu2013semi}. Training-based schemes use long training sequences to obtain the channel state information (CSI) with low complexity but at the expense of decreased throughput \cite{zhang2006optimal}. In contrast, by using the properties of the transmitted signal, blind channel estimation schemes estimate the channel without any training symbols \cite{shin2007blind}. While these schemes are bandwidth efficient, they are more complex and less accurate in comparison with training-based schemes. To strike a balance between throughput, accuracy, and complexity, semi-blind schemes combine the merits of both training-based and blind schemes \cite{murthy2006training}. 
Recently, there has been a growing interest in joint channel estimation and signal detection for grant-free NOMA \cite{jiang2020joint,wang2016joint,wei2016approximate}. By representing the joint problem as a compressed sensing problem, the authors of \cite{chen2018sparse,liu2018massive,jiang2020joint} utilize the sparsity of the pilot (training) symbols to reduce their number. However, in massive IoT, where the packets are usually small, even a few training symbols can lead to a major efficiency loss \cite{salari2020clustering}. %One intuitive way to reduce the number of training symbols with no significant loss of performance is to apply ML algorithms, particularly unsupervised clustering techniques~. 

Unsupervised machine learning algorithms can be used for joint channel estimation and signal detection \cite{salari2022noma}. For an uplink NOMA scenario, the probability density function (PDF) of the received signals can be modelled by a GMM and the maximum-likelihood estimation (MLE) can be used to estimate the model parameters. While the MLE is asymptotically efficient, it may be intractable or biased, especially with a small sample size \cite{rossi2018mathematical,murphy2012machine}. In such cases, the expectation-maximization (EM) algorithm is useful for estimating the model parameters \cite{biship2007pattern,dempster1977maximum} in an iterative manner. EM is numerically stable and increases the likelihood function with each iteration~\cite{meng1993maximum}. The global convergence of the EM algorithm fitting a GMM model consisting of two Gaussian distributions with similar densities is guaranteed, as long as it is initialized in the neighborhood of the desired solution \cite{xu2016global}. For higher-order mixture models, if the GMM centers (the means of the Gaussian components) are well separated, EM converges locally to the global optimum \cite{zhao2020statistical,yan2017convergence}.% with a convergence rate higher than that of the quasi-Newton algorithm~\cite{xu1996convergence}. %However, as shown in~\cite{naim2012convergence}, when there exists small mixing density, the convergence rate will diminish. (unclear sentence)
The lower and upper bounds for the number of samples required to estimate the parameters of a GMM with well-separated centers given any desired accuracy is calculated in \cite{ashtiani2018nearly,kwon2020algorithm}. %Authors of \cite{kwon2020algorithm} showed that for a GMM with $M$ components, EM converges under the cluster separation of $\Omega\left(\sqrt{\log M}\right)$ rather than $\Omega\left(\sqrt{M}\right)$, which was previously presumed \cite{yan2017convergence,zhao2020statistical}.

\subsection{Motivations and Contributions}

In our recent conference paper \cite{salari2020clustering}, the main idea of using clustering algorithms for joint channel estimation and signal detection is presented for an uplink NOMA scenario with quadrature phase shift keying (QPSK) modulation. Using GMM clustering, we can cluster the received signals into $M$ clusters , where $M$ is the modulation order, in each stage of successive interference cancellation (SIC). The channel gain $|h|$ can then be estimated using the coordinates of each cluster centroid. However, this does not provide any information about the phase of the channel. In \cite{salari2020clustering}, we considered sending a few pilot symbols to determine the demapping. However, we did not take into account the pilot length and its impact on the throughput of the system. The results in \cite{salari2020clustering} show that, by using GMM clustering with only a few pilot symbols, the receiver can effectively estimate the users' channels and detect the signals with a SER approaching that of the optimal receiver, that is, the maximum likelihood detection (MLD) with full CSI.

In this paper, we propose a generalized GMM-based joint channel estimation and signal detection algorithm for NOMA. We provide a theoretical analysis of the proposed approach to prove its convergence and derive a closed-form expression for its SER. We also study the tradeoff between pilot length and SER offered by the proposed approach, and further explore its application in a grant-free NOMA scenario suitable for massive IoT systems. Our main contributions in this paper are summarized below.
%
%To better illustrate the advantages of the proposed scheme, we contrasted it to the MLE receiver with full CSI and partial CSI in this paper.     In the first step we have shown that our proposed model requires only 2 pilot symbols to have a performance close to the MLE with full CSI. Next, we demonstrated that in the case of MLE with partial CSI where receiver uses a few pilot symbols to estimate the channel, the MLE receiver requires 8 symbols to produce almost the same SER performance as the suggested GMM clustering approach with only two symbols.
%In this paper, we propose a new approach for joint channel estimation and user detection in uplink NOMA using minimal training symbols and no explicit estimation of the CSI at the receiver. We employ an unsupervised ML algorithm, that is based on a Gaussian mixture model (GMM), to cluster the received signals at the receiver side.
%In this paper, we propose a new algorithm for joint user detection and channel estimation using GMM-based clustering. 

\begin{enumerate}
\item We first propose a generalized GMM-based joint channel estimation and signal detection algorithm for NOMA. We provide a mathematical analysis to characterize the error rate performance of the proposed algorithm. We prove the convergence of the proposed GMM clustering algorithm and derive an upper bound on its error. Accordingly, we calculate a bound on the channel estimation error, in particular, the gain and phase errors, to derive a closed-from expression for the SER. We verify the accuracy of the derived bound via simulations. Furthermore, we show that when the powers of the signals received from the users are sufficiently different, the proposed clustering-based approach with no CSI at the receiver achieves the same SER performance as the conventional MLD receiver with full CSI. We also explore the tradeoff between accuracy and blocklength as the accuracy of clustering algorithm depends on the number of data points (symbols) available at the receiver. We also show that the proposed approach can be easily extended for NOMA with higher-order modulations.%we show that the proposed scheme can closely approach the ML result, when only a few pilot symbols are used for determining the signal demapping. 

%This bound proUsing this bound, we find the In the conference paper~\cite{salari2020clustering}, we only presented the potentials of unsupervised learning (specifically GMM clustering) for combined channel estimation and signal detection. In this paper, Not only did we analyse the performance of this algorithm and the impact of its parameters on the system, but we also developed a mathematical model to evaluate the error rate of the proposed GMM-based joint channel estimation and signal detection technique. Mathematical model results have been verified to match simulation results.

\item We provide a comprehensive comparison between the proposed approach and some existing related ones that use training-based or semi-blind channel estimation. Training-based channel estimation usually requires a large number of pilot symbols to accurately estimate the CSI. We show that our proposed approach considerably outperforms the training-based and semi-blind channel estimation techniques when only a few pilot symbols are available. We also show, the proposed approach requires a significantly lower number of pilot symbols to achieve the benchmark MLD performance compared with existing training-based and semi-blind channel estimation techniques. 

\item We explore the application of the proposed GMM-based joint channel estimation and signal detection algorithm in a practical grant-free NOMA scenario, where the receiver does not have any knowledge of the number of active users or the channel conditions. The results show that the proposed approach almost achieves the MLD performance using a small number of pilot symbols. This is of particular advantage for massive IoT applications, where the users' activity is usually arbitrary and the receiver cannot estimate the channels of all users promptly.

%We demonstrate that our proposed algorithm can be implemented in a real-world situation in which the receiver has no previous knowledge other than the modulation scheme employed by the transmitters and the power levels. As one of the applications of proposed technique, We studied a grant-free situation where the number of active users is unknown to the BS. It has been shown that the proposed GMM-based technique with just two pilot symbols is comparable to MLE with complete CSI in terms of performance.
\end{enumerate}

\subsection{Paper Organization and Notations}

The rest of the paper is organized as follows. We present the system model and provide some relevant preliminary information in Section \ref{System Model}. We describe our proposed GMM-clustering-based approach for joint channel estimation and signal detection in Section \ref{ML}. In Section \ref{Analysis}, we characterize theoretical bounds on the SER of the proposed approach. We provide extensive numerical results to evaluate the performance of the proposed approach in Section \ref{Numerical Results}. The application of the proposed approach in grant-free massive IoT is presented in Section \ref{Grant-free}, Finally, Section~\ref{Conclusion} concludes the paper. 

\emph{Notations:}
Matrices, vectors, and scalars are denoted by uppercase boldface, lowercase boldface, and lowercase letters, e.g., $\mathbf{A}$, $\mathbf{a}$, and $a$, respectively.
The transpose, inverse and norm of $\mathbf{A}$ are denoted by $\mathbf{A}^T$ , $\mathbf{A}^{-1}$, and $||\mathbf{A}||$, respectively.
% The $\ell_2$-norm of $\mathbf{a}$ and the Frobenius norm of $\mathbf{A}$ are denoted by $||\mathbf{a}||$ and $||\mathbf{A}||_{F}$, respectively.
%
%The $N \times N$ identity matrix and the $N \times N$ matrix consisting of all zero entries are denoted by $\mathbf{I}_N$ and $\mathbf{0}_N$, respectively.
%The operators $\mathbb{E}[.]$ , tr(.), det(.), vec(.), mat(.), $\otimes$, and $\prod$ indicate the expectation, trace of a matrix, determinant of a matrix, matrix vectorization, inverse operation of vec(.), Kronecker product, and sequence product operators, respectively.

%
%
%
%
\section{System Model and Preliminaries}\label{System Model}
%
%This section begins with a definition of the channel and signal model, followed by a quick overview of the fundamentals of GMM-based clustering.
%
%
\subsection{Channel Model}

We consider a cellular uplink NOMA scenario in which $K$ active users simultaneously transmit packets of length $N$ symbols to a base station (BS).  
Let $x_i$ denote the signal transmitted by user $i$, which is drawn from the signal constellation $\mathcal{S}=\{\mathbf{s}_1,\mathbf{s}_2,\cdots,\mathbf{s}_M\}$ with $M=|\mathcal{S}|$ being the modulation order. The received signal at the BS, denoted by $y$, is given by
\begin{align} \label{eq:1}
    y=\sum_{i=1}^{K}h_i\sqrt{P_i}x_i+w,
\end{align}
where $P_i$ is the transmit power of user $i$ and $w\sim \mathcal{CN}(0,1)$ is additive white Gaussian noise (AWGN). The gain of the channel between user $i$ and the BS is denoted by $h_i$, which includes both small-scale and large-scale fadings, i.e., $h_i=g_i\sqrt{\ell_0(r_i/r_0)^{-\alpha}\chi}$, where $\ell_0$ and $r_0$ are the reference path-loss and reference distance, respectively, $\alpha$ is the path-loss exponent, $r_i$ is the distance between user $i$ and the BS, $\chi$ is the large-scale shadowing modeled by a log-normal distribution with zero mean and variance $\sigma$ dB, and $g_i$ is the small-scale fading modelled by the Rayleigh, Rician, Nakagami, or any other distribution. The signal-to-noise ratio (SNR) of user $i$ at the BS is given by $\gamma_i=P_i|h_i|^2$. We assume that $h_i$ remains constant for the duration of one packet ($N$ symbols), which is a valid assumption for short packets, especially in mMTC applications~\cite{yu2020joint,lv2021energy}.

Given the user channel gains and knowing that all users utilize the same modulation with constellation set $\mathcal{S}$, the probability distribution of the received signals at the BS can be expressed as a mixture of Gaussian distributions\footnote{One can easily show that this is also valid when users use different modulations, and the assumption is made only for a better representation in the paper.}. Let $\mathbf{h}=[h_1,h_2,\cdots,h_K]$,  $\mathbf{p}=[P_1,P_2,\cdots,P_K]$, and $\mathbf{x}=[x_1,x_2,\cdots,x_K]$ denote the vectors of channel gains, transmit powers, and user signals, respectively. Therefore, we have%\mathbf{u}_1\in\mathcal{S},\cdots,\mathbf{u}_K\in\mathcal{S}
\begin{align}
    p\left(y|\mathbf{h},\mathbf{p},\mathcal{S}\right)&= \sum_{\mathbf{u}\in\mathcal{S}^K} p(\mathbf{x}=\mathbf{u}) p(y|\mathbf{h},\mathbf{p},\mathcal{S},\mathbf{x}=\mathbf{u})\\
    & =\sum_{\mathbf{u}\in\mathcal{S}^K}\prod_{i=1}^K p(x_i=u_i) p(y|\mathbf{h},\mathbf{p},\mathcal{S},\mathbf{x}=\mathbf{u})\\
    &=\frac{1}{|\mathcal{S}|^K}\sum_{\mathbf{u}\in\mathcal{S}^K}p\left(\left.\sum_{i=1}^K \sqrt{P_i}h_iu_i+w\right|\mathbf{h},\mathbf{p}\right), \label{eq:mixture of Gaussian}
\end{align}
where we assume that the signals are randomly drawn from $\mathcal{S}$ with equal probabilities, i.e., $p(x_i=u_i)=1/|\mathcal{S}|$. Since $w\sim\mathcal{CN}(0,1)$, we have $\sum_{i=1}^K\sqrt{P_i}h_iu_i+w\sim\mathcal{CN}(\sum_{i=1}^K\sqrt{P_i}h_iu_i,1)$. Therefore, we can simplify \eqref{eq:mixture of Gaussian} as follows
\begin{align} \label{eq:GMM justification}
    y|\mathbf{h}\sim \frac{1}{|\mathcal{S}|^K}\sum_{\mathbf{u}\in{\mathcal{S}}^{K}}\mathcal{CN}\left(\sum_{i=1}^K\sqrt{P_i}h_iu_i,~ 1\right),
\end{align}
which means that the distribution of the received signal follows a Gaussian mixture model.

We assume that all users are frame-synchronized, which is achieved by frequently sending beacon signals from the BS~\cite{hasan2018time}. We also assume that the BS does not know the CSI for any user. Therefore, it attempts to jointly estimate the channels and detect users' signals. However, we assume that the BS knows the number of transmitting users, $K$, and the utilized modulation scheme. Later in Section~\ref{Grant-free}, we will show how the assumption of knowing the number of users can be relaxed. %\textcolor{blue}{(knowledge of $K$ natural in grant-based systems, but needs to be supported by references if grant-free is considered)}.
\subsection{GMM Clustering}
%Since channels are block-fading and all users employ the same modulation scheme, the clusters at the BS are symmetric in the I-Q plane with roughly the same densities (see Fig. \ref{fig:constellation}). 
%
\begin{figure}[t]
  \begin{center}
    \includegraphics*[width=0.9\columnwidth]{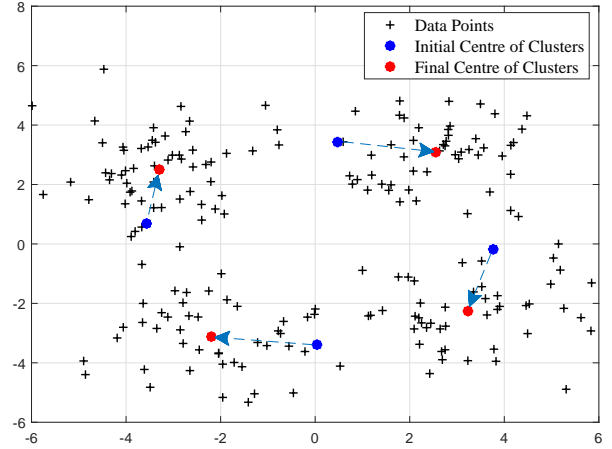}
     \caption{\small{EM mean convergence after eight iterations for a single-user scenario when $\gamma = 7$dB and $N=200$.}}
    \label{fig:Effect_of_iteration}
  \end{center}
\end{figure}

Fig. \ref{fig:Effect_of_iteration} shows the signals received at the BS for the single user scenario employing the QPSK modulation and blocklength of $N=200$. As seen, the received signals can be effectively clustered into four clusters. The GMM clustering algorithm \cite{singh2009statistical} finds the centroid of these clusters in an iterative manner using the EM algorithm.

Given that the received signals are complex numbers, we denote a $2$-dimensional multivariate Gaussian probability density function by $g(\mathbf{z};\bm{\mu},\bm{\Sigma})$ where $\bm{\mu}$ and $\mathbf{\Sigma}$ are the mean vector and the covariance matrix, respectively, and express it as
\begin{equation}\label{eq:3}
      g(\mathbf{z};\bm{\mu},\bm{\Sigma}) = \frac{\exp{ \left(-\frac{1}{2} (\mathbf{z} -\bm{ \mu})^T \mathbf{\Sigma}^{-1} (\mathbf{z} - \bm{\mu})  \right)}}{\sqrt{(2\pi)^{2} |\mathbf{\Sigma}|}}.  
\end{equation}

In GMM clustering, the number of clusters is pre-determined and the data is assumed to be generated by a mixture of Gaussian distributions. A GMM parameterizes the mean, covariance, and weight of each Gaussian distribution component. When a common $M$-ary modulation scheme is adopted by all users, there are $M$ Gaussian distributions each with a nonnegative mixture weight $\omega_j$ where $j\in\{1,\cdots,M\}$ and $\sum_{j=1}^M\omega_j=1$. Accordingly, the underlying Gaussian mixture distribution can be written as a convex combination of $M$ constituent Gaussian distributions (each representing a cluster), i.e.,
\begin{equation}\label{eq:4}
       p(\mathbf{z};\bm{\mu}_1,...,\bm{\mu}_M, \bm{\Sigma}_1,...\bm{\Sigma}_{M}) = \sum_{j=1}^{M} \omega_j g_j(\mathbf{z};\bm{\mu}_j,\mathbf{\Sigma}_j).
 \end{equation}
We are interested in estimating $\bm{\mu}_j$, $\mathbf{\Sigma}_j$, and $\omega_j$, $j=1,\cdots, M$, from the observed data. This can be done by maximizing the likelihood function \eqref{eq:4} for all received signals. To this end, we utilize the EM algorithm~\cite{hastie2009elements} that is suitable for solving maximum-likelihood problems containing unobserved latent variables.

We define the corresponding log-likelihood function as
\begin{align}\label{eq:6}
       l^{(t)} (\bm{\mu}_1,&\cdots,\bm{\mu}_{M}, \bm{\Sigma}_1,\cdots,\bm{\Sigma}_{M}|\mathbf{z}_1,\cdots,\mathbf{z}_N) \\ \nonumber
       & = \sum_{i=1}^{N} \left[ \sum_{j=1}^{M} \Delta_{i,j}^{(t)} \ln \left( \omega_j^{(t)} g_j\left(\mathbf{z}_i;\bm{\mu}_j^{(t)},\mathbf{\Sigma}_j^{(t)}\right) \right)  \right].
 \end{align}
We evaluate this function only to verify the convergence of the EM algorithm. Let $\Delta_{i,j}$ symbolize the association of the $i$th data point to the $j$th cluster represented by the $j$th Gaussian distribution. Therefore, we have
$$ \Delta_{i,j}= \begin{cases} 1; \hspace{5mm} \text{if} \hspace{1mm} \mathbf{z}_i \hspace{1mm} \text{belongs to cluster} \hspace{1mm} g_j,\\
0; \hspace{5mm} \text{otherwise}. \end{cases} $$
It is clear that $P(\Delta_{i,j} = 1)  = \omega_j$ and $P(\Delta_{i,j} = 0)  = 1 - \omega_j$. However, both $\Delta_{i,j}$ and $\omega_j$ are unknown. 

In the $t$th iteration of the EM algorithm, we first estimate the so-called responsibility variable for each model $j$ and every data point $i$ as
\begin{align}\label{eq:5}
        \hat{\gamma}_{i,j}^{(t)} = \frac{\hat{\omega_j}^{(t-1)} g_j\left(\mathbf{z}_i;\hat{\bm{\mu}}_j^{(t-1)}, \hat{\mathbf{\Sigma}}_j^{(t-1)}\right)}{\sum_{k=1}^{{M}} \hat{\omega_{k}}^{(t-1)} g_{k}\left(\mathbf{z}_i;\hat{\bm{\mu}}_{k}^{(t-1)}, \hat{\mathbf{\Sigma}}_{k}^{(t-1)}\right)}.
 \end{align}
We then assign each data point to its corresponding cluster. In particular, for each $\mathbf{z}_i$, we find $m^{(t)}_i=\arg\max_j\hat{\gamma}^{(t)}_{i,j}$ and set 
$$ \Delta^{(t)}_{i,j}= \begin{cases} 1; \hspace{5mm} \text{if} \hspace{1mm} j=m^{(t)}_i,\\
0; \hspace{5mm} \text{otherwise}. \end{cases} $$
In the next step of the EM algorithm, we use the calculated responsibilities to update the mixture weight, mean, and variance for each cluster as follows
\begin{align}\label{eq:7}
       \hat{\omega}_j^{(t)}  &= \frac{ \sum_{i=1}^{N} {\hat{\gamma}}_{i,j}^{(t)}}{ \sum_{i=1}^{N} \sum_{k=1}^{{M}}  {\hat{\gamma}}_{i,k}^{(t)} }, \\
\label{eq:8}
       \hat{\bm{\mu}}_j^{(t)}  &= \frac{ \sum_{i=1}^{N} \hat{\gamma}_{i,j}^{(t)} \mathbf{z}_i }{ \sum_{i=1}^{N}  \hat{\gamma}_{i,j}^{(t)} },
\\
\label{eq:9}
       \hat{\mathbf{\Sigma}}_j^{(t)}  &= \frac{ \sum_{i=1}^{N} \hat{\gamma}_{i,j}^{(t)} \left[\mathbf{z}_i - \hat{\bm{\mu}}_j^{(t)}\right] \left[\mathbf{z}_i - \hat{\bm{\mu}}_j^{(t)}\right]^T }{ \sum_{i=1}^{N}  \hat{\gamma}_{i,j}^{(t)}. }. 
\end{align}

After enough iterations, the values of responsibility, mean, covariance, and mixture weight for each cluster converge as the EM algorithm is guaranteed to converge to a local optimum~\cite{dempster1977maximum}. The number of iterations required for convergence mainly depends on the convergence criterion. In Fig.~\ref{fig:Effect_of_iteration}, we visualize the convergence of the mean estimates (cluster centroids) after running eight iterations of the EM algorithm.
%We have observed that the algorithm usually converges in less than ten iterations.

There are several other prominent clustering algorithms, such as $k$-means, DBSCAN~\cite{ester1996density}, OPTICS~\cite{ankerst1999optics}, and mean shift~\cite{cheng1995mean}, that can also be used to cluster the received signals. However, as the noise has Gaussian distribution, the received signals at the BS naturally follow a mixture of Gaussian distributions~\eqref{eq:GMM justification}. Hence, GMM is a good choice for our clustering problem. 
From a theoretical standpoint, given that the noise affecting the received signals is independent and identically distributed (i.i.d.) AWGN, the GMM clustering used in our proposed algorithm is equivalent to the well-known $k$-means clustering algorithm. However, in practise, due to the limited amount of observations, the noise covariance matrix is not a multiple of the identity matrix, which means that the noise affecting different received signals may be correlated or have different variances. As a result, GMM clustering is more accurate compared to $k$-means clustering since, unlike $k$-means, it does not assume the same covariance for all clusters but estimates them from the data.

\section{Clustering-based Joint Channel Estimation and Signal Detection}  
\label{ML} 

\begin{figure}[t]
  \begin{center}
    \includegraphics*[width=1\columnwidth]{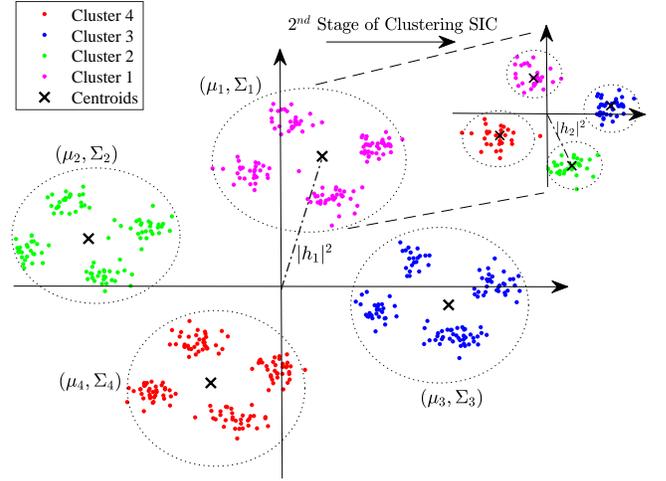}
     \caption{\small{Received signal constellation diagram of a two-user NOMA system at the BS for $N=500$,  $\gamma_1 =17$dB, and $\gamma_2 =11$dB.}}
    \label{fig:constellation}
  \end{center}
\end{figure}

To better understand the proposed joint channel estimation and signal detection approach, we first discuss Fig.~\ref{fig:constellation} that illustrates the signals collected at the receiver in the I-Q plane for a two-user NOMA communication system when both users employ the QPSK modulation.
%Fig. \ref{fig:constellation} show the received signals for a two-user NOMA communication system, when both users use the QPSK modulation. As seen in Fig.\ref{F1a}, for a single active user, the received signals at the BS can be grouped into four clusters and the centroid of each cluster can be used to estimate the amplitude and phase of the channel. When the number of active users is increased by one, the signals of both users can be grouped into 16 clusters as the BS knows the number of active users (Fig. \ref{F1b}). 
We assume that $|h_1|\ge |h_2|$ without losing generality. Fading causes signal amplification and a rotation with respect to the original constellation. When the received powers for the two users are significantly different, the clusters are distinct from each other. Therefore, the phase and amplitude of the channels can be estimated accurately. However, when channel fading and noise cause the clusters to overlap, estimation of the channels and detection of the signals is more challenging and inevitably less accurate. In what follows, we propose an effective algorithm to jointly estimate the user channels by clustering the received signals and detect the user signals.

\begin{algorithm}[t]
        \KwIn{number of users $K$, received signal $\mathbf{y}$, modulation order $M$, signal constellation $\mathcal{S}$, and convergence threshold $\epsilon$}
        \KwOut{estimated signals of the users}

            Set $\omega_j = \frac{1}{M}, ~~j=1,\cdots,M$

         \For{ user $u= 1$ to $K$  }
           {     
                \textbf{Initialize} $\hat{\bm{\mu}}^{(0)}$ and $\hat{\mathbf{\Sigma}}^{(0)}$ by dividing the received signals on the reference coordinate system into $M$ equal sections, select one point in each section as the initial mean and set the initial covariance to one
                
                Calculate $\hat{\gamma}_{i,j}^{(0)}$ according to (\ref{eq:5})
                
                Calculate log-likelihood function according to (\ref{eq:6})
                
                Set $t=1$
               
            \While {$ l^{(t)} - l^{(t-1)} \geq \epsilon $}
                {
                Update $\hat{\bm{\mu}}^{(t)}$ and $\hat{\mathbf{\Sigma}}^{(t)}$ using (\ref{eq:8}) and (\ref{eq:9})
                
                Update $\hat{\gamma}_{i,j}^{(t)}$ according to (\ref{eq:5})
                
                Update log-likelihood function according to (\ref{eq:6})
                }
            \textbf{Return} $\hat{\bm{\mu}}$=$\hat{\bm{\mu}}^{(t)}$ and $\hat{\mathbf{\Sigma}}=\hat{\mathbf{\Sigma}}^{(t)}$
            
            Calculate the phase of each cluster centroid: $\phi_i = \tan^{-1}\left(\frac{\mathrm{Im}(\hat{\mathbf{\mu}_i})}{\mathrm{Re}(\hat{\mathbf{\mu}_i})} \right)$
            
            Calculate the average phase as  $\phi = \frac{\sum_{i=1}^{M} {\phi_i} - M \pi}{M} $ and channel amplitude $|\hat{h}_u|=\frac{1}{M}\sum_{i=1}^M \mathrm{abs}(\mu_i)$
            
            Update the decision boundaries based on the $\phi$ and use the pilot symbol to map each cluster into the mapping bits
            
            Demodulate the signal to symbols: $\hat{\mathbf{x}}_u=\mathrm{demod}(\mathbf{y})$
            
            %Estimate the channel phase rotation using the pilot symbols and \eqref{eq:MMSE}
            
            Re-modulate this user's signal and multiply by the estimated channel gain and subtract it from superimposed received signal: 
            $\mathbf{y}\leftarrow\mathbf{y}-|\hat{h}_u|e^{j\phi} \hat{\mathbf{x}}_u$
        }
        \caption{Joint channel estimation and signal detection using GMM clustering.}
        \label{GMM Clustering Algorithm}
        %\vspace{-1em}
\end{algorithm}

At the BS, SIC is used to recover the multiplexed user signals from the received superimposed signals in the decreasing order of their received powers. Initially, the receiver detects the signal of the strongest user that is the user with the highest received power. It then reconstructs and removes it from the received signal. Afterwards, it detects the next strongest signal and so on. In order to implement SIC at the BS, we start by dividing the received signals (data points) into $M$ (four with QPSK) clusters, which represent the user 1's signals. Next, we further split each of the $M$ clusters into $M$ smaller clusters representing the user 2's signals and so on. Taking the received signals as our observed data, the considered joint channel estimation and signal detection problem boils down to estimating the unknown latent parameters of the assumed Gaussian mixture distribution in~\eqref{eq:GMM justification}. 

%When the signals of the users are uniformly drawn from the same QPSK constellation, the weights of the Gaussian distributions are the same, i.e., $\omega_j = \frac{1}{4}$, $j=1,\cdots,4$. 
\subsection{The Proposed Algorithm}

We summarize the proposed algorithm for joint channel estimation and signal detection in Algorithm \ref{GMM Clustering Algorithm}. In the proposed algorithm, at each stage of SIC, we estimate the parameters of only $M$ Gaussian distributions. This helps with managing the computational complexity.

\begin{comment}
\begin{figure*}[t] 
\subfloat[Single user, $\gamma = 12$dB \label{F1a}]{%
\includegraphics[width=0.667\columnwidth]{Constellation-SingleUser.eps}
}
\hfill
    \subfloat[2-user NOMA, $\gamma_1 =20$dB, $\gamma_2 =17$dB   \label{F1b}]{%
\includegraphics[width=0.666\columnwidth]{Constellation-NOMA-20-17.eps}
}
\hfill
\subfloat[2-user NOMA, $\gamma_1 =10$dB, $\gamma_2 =7$dB  \label{F1c}]{%
\includegraphics[width=0.666\columnwidth]{Constellation-NOMA-10-7.eps}
}
\caption{\small{Received signal constellation diagram at the BS for $N=500$ and different values of SNR ($\gamma_u$).}}
\label{fig:constellation}
\vspace{-1.5em}
\end{figure*}
\end{comment}
%%%%%%%%%%%%%%%%
%
%
%

Given the modulation order $M$, we fix the mixture weight of each cluster as $\omega_j = \frac{1}{M}$. Next, starting from the first user, we split the received data into $M$ clusters and select one point in each cluster as the initial mean and set the initial variance of each cluster to one.
Then, we calculate the responsibility and log-likelihood function values using \eqref{eq:5} and \eqref{eq:6}, respectively. Afterwards, we calculate the associated cluster centroids and covariance matrices (lines 8 to 13 in Algorithm \ref{GMM Clustering Algorithm}). We continue by evaluating the phase of each cluster. Considering that we are using the QPSK modulation, the phase difference between any two adjacent clusters is $\frac{\pi}{2}$. However, due to noise, the phase of each centroid might differ from $\frac{\pi}{4}, \frac{3\pi}{4}, \frac{5\pi}{4}$, or $\frac{7\pi}{4}$. To minimize the effect of phase rotation, we average the phase difference between the centroid of each cluster and its expected value (line 14 of Algorithm \ref{GMM Clustering Algorithm}), and update the decision boundaries according to the average phase rotation. Finally, we apply SIC and repeat the algorithm for the next strongest user.

It is important to note that by using SIC, we assume the first $M$ clusters to be Gaussian each consisting of $M$ subclusters, although, the clusters corresponding to the strongest user are not strictly Gaussian in practice. However, roughly speaking, the components of the GMM represent clusters that are of similar shapes. Thus, each cluster is implicitly assumed to follow a Gaussian distribution. For cases where this assumption is overly unrealistic, a natural alternative is to assume that each cluster is also a mixture of normally-distributed subclusters~\cite{hastie1996discriminant}.

\subsection{Symbol-to-Bit Demapping}

\begin{figure}[t]
    \centering
    \includegraphics[width=0.9\columnwidth]{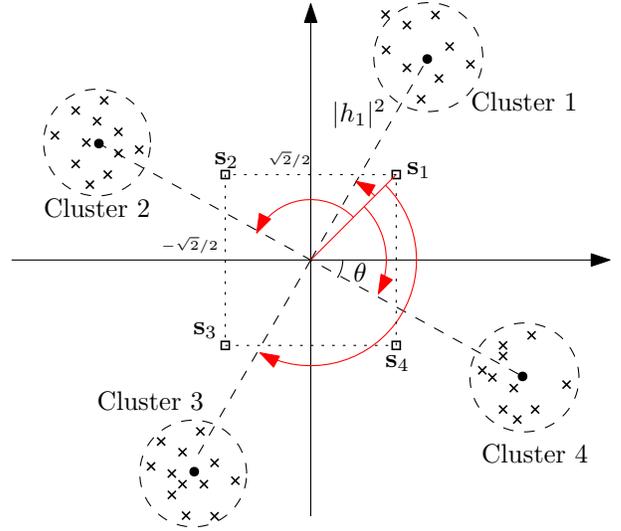}
    \caption{Constellation rotation due to channel fading.}
    \label{fig:singleuserE}
\end{figure}

Using GMM clustering, we can separate the received signals into $M$ clusters in each stage of SIC. We can then estimate the channel gain $|h|$ from the coordinates of the cluster centroids (line 14 of Algorithm \ref{GMM Clustering Algorithm}). However, this does not inform us about the exact phase of the channel. In particular, assuming that the user sends QPSK symbols, according to Fig. \ref{fig:singleuserE}, one can consider four different but equally probable choices for the phase of the channel, i.e., $\angle h\in\{\theta-\pi/4, \theta-3\pi/4,\theta+\pi/4, \theta+3\pi/4\}$ where $-\pi/4\le\theta\le\pi/4$ is the phase of the cluster that resides in the angular region of $[-\pi/4,\pi/4]$. This implies that we cannot demap the signals to the symbols. 

To overcome this issue, we send a few pilot symbols to determine the demapping. Determining the mapping with only one symbol is unambiguous when the SNR is not excessively low. For a robust and accurate performance, two symbols per user can be sufficient. We use a minimum mean square error (MMSE) method to estimate the phase rotation due to the channel. The MMSE estimate is given as
\begin{equation} \label{eq:MMSE}
\hat{h}_{u,\textrm{MMSE}} = \left({x^H_{u,p}}{x_{u,p}}+1\right)^{-1}{x^H_{u,p}}{y_{u,p}}
\end{equation} 
where $x_{u,p}$ and $y_{u,p}$ are the transmitted and received pilot symbols, respectively. The phase of the channel is determined to be the one from $\{\theta-\pi/4, \theta-3\pi/4,\theta+\pi/4, \theta+3\pi/4\}$ that is the closest to $\hat{h}_{u,\textrm{MMSE}}$. Using this method, the exact phase rotation due to the channel is not required. As we will show later, this is another advantage of our proposed approach over the conventional channel estimation techniques. 
Referring to Fig.~\ref{fig:singleuserE}, for two-user NOMA, we may assume that, in the first time slot, both users send $s_1$ while, in the second time slot, the first user transmits $s_3$ and the second user sends $s_1$ as the pilot symbols.
%\hl{how do the pilot symbols look like?}

\section{Theoretical Analysis}\label{Analysis}

%\hl{This is really confusing, please consider rewriting.}
%In the previous section, we implemented GMM to cluster the received data, and the detection has been performed at the receiver accordingly. Since optimizing the likelihood function is hard, we utilized the EM algorithm to estimate the GMM parameters. Then, we have evaluated the SER at the receiver based on the number of received data points. However, as it has been shown, to evaluate the SER, we need to run the proposed algorithm. In the following, we will propose a method to characterize the SER based on the parameters of systems. 
%
%So far, we have proposed an algorithm based on GMM clustering for joint channel estimation and signal detection. Since optimizing the likelihood function is hard, we utilized the EM algorithm to estimate the GMM parameters.
In this section, we analyze the performance of the proposed algorithm. In each iteration of EM, the mean, variance, and mixture weight of each cluster are updated. We refer to the difference between the mean of each cluster estimated by EM and its corresponding exact value as the EM error. We start by presenting a theorem that gives an upper-bound on the EM error. Then, using this bound, we present a mathematical model to predict the SER of our proposed GMM-clustering-based joint channel estimation and signal detection algorithm. 
%
%
%\subsection{Evaluating the Error of EM}\label{}
%
%\hl{Please consider rewriting and avoid general terms like data, etc. You are talking about the signals at the receiver and you should be consistent in the entire paper.}

\begin{figure}[t]
  \begin{center}
    \includegraphics[width=0.9\columnwidth]{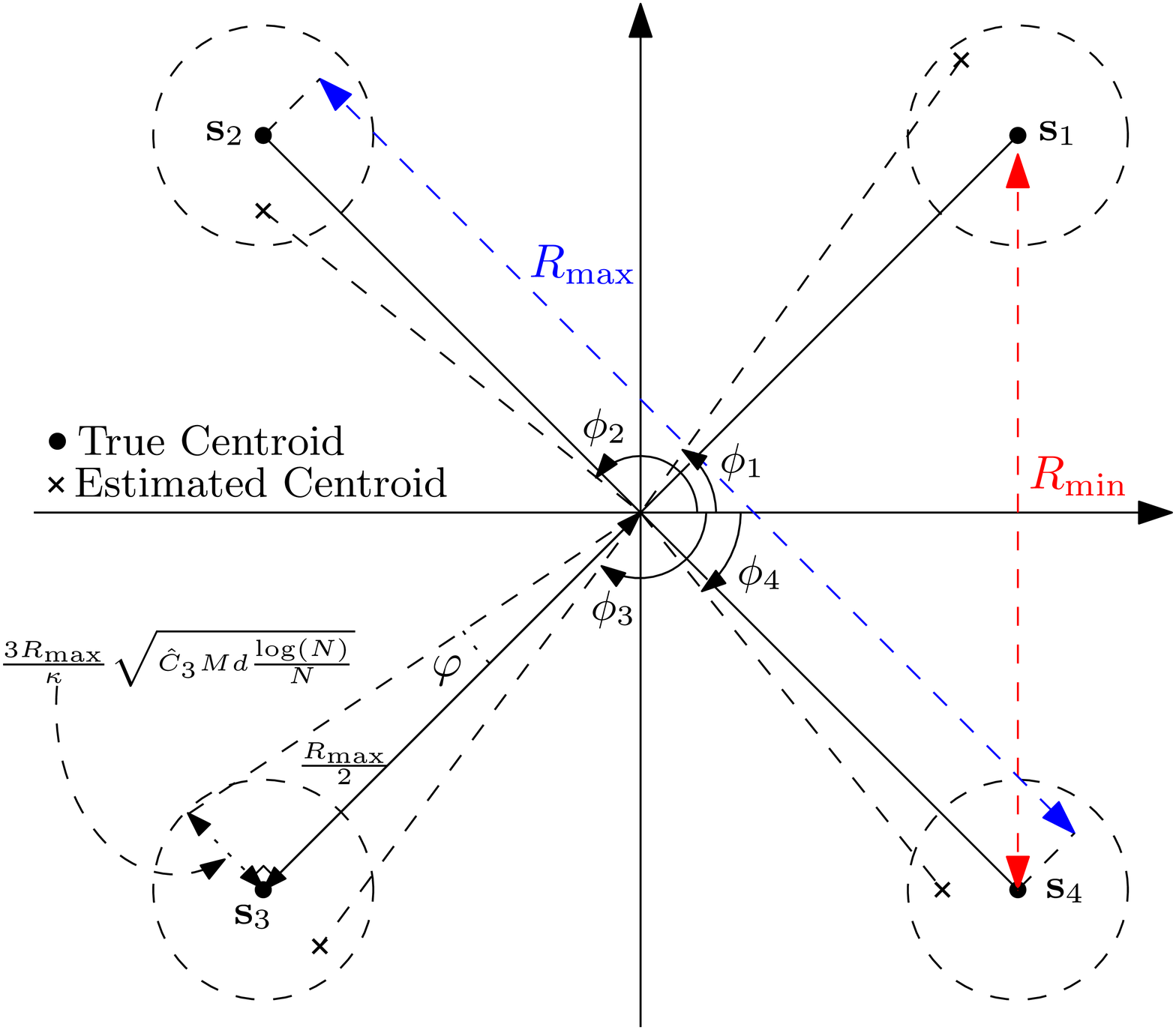}
  \caption{\small{The convergence of the EM algorithm when $M=4$ and $d=2$.}}
  \label{fig:arbitrary_rotation}
  \end{center}
  \vspace{-1em}
\end{figure}

Let $R_{\min}$ and $R_{\max}$ denote the distance between the closest cluster centroids and the distance between the farthest cluster centroids, respectively, as show in Fig. \ref{fig:arbitrary_rotation}. The following theorem characterizes an upper-bound on the EM error. %We also denote the Referring to Fig. \ref{fig:Effect_of_iteration}, we consider the distance between the red dots (final center of the clusters) as $R$, and we denote the distance between the closest red dots as $R_{\min}$ and the distance between the farthest red dots as $R_{\max}$. 

%(For QPSK, $R=2 \sqrt{ \varepsilon_b }$  since minimum distance in QPSK is $2\sqrt{(\textrm{log}_2 M \times \sin^2(\frac{\pi}{M})) \varepsilon_b }$ where $\varepsilon_b $ is the bit energy). 
%\hl{I suggest adding a figure to clearly demonstrate what you mean.}

%\hl{The theorem should be more detailed and should include all things required to understand it. Please consider rewriting. For example you need to define $\mu_i$, $M$, etc. Use $\min$ instead of $min$ and $\max$ instead of $max$.}
\begin{thm} \label{thm:1}
Assume $d$-dimensional received signals and $M$ isotropic Gaussian distributions with mixture weights $\omega_j$, $j\in\{1,\cdots,M\}$. Let $\bm{\mu}_i^{*}$ denote the true mean of cluster $i$ and $\bm{\mu}_i^{(t)}$ represent the estimated mean of the $i$-th cluster after $t$ iterations of EM. Suppose $\kappa=\min\{ \omega_j \}$, $R_{\min} \geq C_0 \sqrt{\min \{d,M\}}$, and the initial iterate $\bm{\mu}^{(0)}$ satisfies
\begin{align}
    \nonumber &\underset{ i \in [M] }{\max} \hspace{2mm}|| \bm{\mu}_i^{(0)} - \bm{\mu}_i^{*} ||_2\leq \frac{R_{\min}}{2} \\
    & - \frac{C_1}{2} \sqrt{\min \{d,M\}}
    \log \left(\max\left\{ \frac{M}{\kappa^2}, R_{\max}, \min\{d,M\} \right\}  \right)
    \end{align} 
where $[M]=\{1,2,\cdots, M\}$ and $C_0,C_1>0$ are universal constants. For a sufficiently large sample size $N$ such that that
\begin{equation} \label{eq:sample_constraint}
    \frac{\log (N)}{N} \leq \min \left\{ \frac{\kappa^2}{144 \hat{C_2} M d},\frac{\kappa^2 \underset{ i \in [M] }{\max} \hspace{1mm}|| \bm{\mu}_i^{(0)} - \bm{\mu}_i^{*} ||_2^2 }{9 \hat{C_3} R_{\max}^2 M d} \right\}
\end{equation} 
where $C_2,C_3 > 0$ are universal constants and $$\hat{C_2} = C_2 \log \left( M\left( 2R_{\max} + \sqrt{d} \right)  \right)$$ $$\hat{C_3} = C_3 \log \left( M\left( 3 R_{\max}^2 + \sqrt{d} \right)\right),$$
the subsequent EM iterates $\{\bm{\mu}_i^{(t)}\}^{\infty}_{t=1}$ satisfy
\begin{align} \label{eq:thm_equation}
    \underset{ i \in [M] }{\max} \hspace{1mm}|| \bm{\mu}_i^{(t)} - \bm{\mu}_i^{*} ||_2  \leq & \frac{1}{2^t} \underset{ i \in [M] }{\max} \hspace{1mm}|| \bm{\mu}_i^{(0)} - \bm{\mu}_i^{*} ||_2 \nonumber \\
    & + \frac{3 R_{\max}}{\kappa} \sqrt{ \frac{\hat{C_3} M d \hspace{1mm} \log (N) }{N}}
\end{align} 
with probability at least $1-\tfrac{2M}{N}$. 
\end{thm}

\begin{IEEEproof}
See Appendix A. 
\end{IEEEproof}

Theorem \ref{thm:1} states that the EM error is bounded. As the iteration number, $t$, increases, the first term on the right-hand side of \eqref{eq:thm_equation} converges to zero and the second term dominates. The following remark characterizes this bound.

\begin{rem}
\label{rem1}
When $t\to \infty$, the EM error characterized in Theorem \ref{thm:1} is upper bounded by
\begin{align} \label{eq:rem1}
    \underset{ i \in [M] }{\max} \hspace{1mm}|| \bm{\mu}_i^{(\infty)} - \bm{\mu}_i^{*} ||_2  \leq \frac{3 R_{\max}}{\kappa} \sqrt{ \frac{\hat{C_3} M d \hspace{1mm} \log (N) }{N}}.
\end{align} 
\end{rem}

This shows that EM converges to points within balls of radius $\frac{3 R_{\max}}{\kappa} \sqrt{ \frac{\hat{C_3} M d \hspace{1mm} \log (N) }{N}}$ around the true centers. Fig.~\ref{fig:arbitrary_rotation} illustrates the essence of Theorem \ref{thm:1} when $M=4$. As seen in the figure, there is a circle of radius $\frac{3 R_{\max}}{\kappa} \sqrt{ \frac{\hat{C_3} M d \hspace{1mm} \log (N) }{N}}$ around each true center and the EM converges to points within these circles. Moreover, the EM error converges to zero when the sample size, $N$, is arbitrarily large. This follows directly from \eqref{eq:rem1} by taking $N$ to infinity, noting that $\hat{C}_3$ is constant.

\begin{rem}
\label{rem2}
The EM error characterized in Theorem \ref{thm:1} converges to zero when $t\to \infty$ and $N\to \infty$.
\end{rem}

The following lemma provides an approximation for the SER of the proposed GMM-clustering-based joint channel estimation and signal detection algorithm when $M=4$, i.e., when QPSK modulation is used.

\begin{lem}
\label{lem1}
For a QPSK-modulated signal, i.e., $M=4$ and $d=2$, at SNR $\gamma$ with $N$ samples, the proposed algorithm achieves an SER approximated by
\begin{align}
P_{\mathrm{SER}}\approx Q\left(\sqrt{2\gamma}\sin\left(\frac{\pi}{4}-\varphi\right)\right)+Q\left(\sqrt{2\gamma}\sin\left(\frac{\pi}{4}+\varphi\right)\right)
\label{eq:serapprox}
\end{align}
where 
\begin{align}
    \varphi=\tan^{-1}\left(6M\sqrt{\frac{\hat{C}_3Md\log(N)}{N}}\right)
    \label{eqphase}
\end{align}
and $Q(z)=\frac{1}{\sqrt{2\pi}}\int_z^{\infty}e^{-\frac{\nu^2}{2}}d\nu$.
\end{lem}

\begin{IEEEproof}
Considering Remark~\ref{rem1}, each estimated cluster centroid has a distance of at most $\frac{3 R_{\max}}{\kappa} \sqrt{ \frac{\hat{C_3} M d \hspace{1mm} \log (N) }{N}}$ to the true centroid when $t\to\infty$. This means for QPSK-modulated signals ($M=4$), as shown in Fig.~\ref{fig:arbitrary_rotation}, there is a phase mismatch between the final cluster centroids and their corresponding true centroids given by
\begin{align}
    \nonumber\varphi&=\tan^{-1}\left(\frac{\frac{3R_{\max}}{\kappa}\sqrt{\frac{\hat{C}_3Md\log(N)}{N}}}{\frac{R_{\max}}{2}}\right)\\
    &=\tan^{-1}\left(6M\sqrt{\frac{\hat{C}_3Md\log(N)}{N}}\right)
\end{align}
where we use $\kappa=1/M$. Therefore, the detection of the QPSK signals is with a rotated phase reference of $\varphi$. In~\cite{some1995bit}, it is shown that the SER for a QPSK signal with a noisy phase reference $-\frac{\pi}{4}\le \varphi\le \frac{\pi}{4}$ is given by \eqref{eq:serapprox}. This completes the proof. 
\end{IEEEproof}

Remark \ref{rem2} states that, when the number of samples $N$, i.e., the number of signals received at the BS, is sufficiently large, the EM error converge to zero. This means that the GMM-clustering-based joint channel estimation and signal detection algorithm will find the true locations of the cluster centroids (shown by black dots in Fig.~\ref{fig:arbitrary_rotation}). Thus, the mismatch in the phase reference $\varphi$ will be zero (Lemma \ref{lem1}) and $P_{\mathrm{SER}}\approx2Q(\sqrt{\gamma})$. Therefore, the SER of the proposed algorithm will be the same as that of the MLD with full CSI. 

We can further extend the results in Lemma~\ref{lem1} to approximate the SER of the proposed algorithm for the two-user NOMA scenario.

\begin{lem}
\label{lem2}
For a two-user NOMA scenario with $N$ QPSK-modulated signals where the SNR of users 1 and 2 are $\gamma_1$ and $\gamma_2$, respectively, the proposed algorithm yields per-user SERs approximated by
\begin{align}
\nonumber P_{\mathrm{SER},1}&\approx \frac{1}{4}Q\left(\left(\sqrt{2\gamma_1}+\sqrt{2\gamma}_2\right)\sin\left(\frac{\pi}{4}-\varphi_1\right)\right)\\
\nonumber &+\frac{1}{4}Q\left(\left(\sqrt{2\gamma_1}+\sqrt{2\gamma_2}\right)\sin\left(\frac{\pi}{4}+\varphi_1\right)\right)\\
\nonumber &+\frac{1}{4}Q\left(\left(\sqrt{2\gamma_1}-\sqrt{2\gamma}_2\right)\sin\left(\frac{\pi}{4}-\varphi_1\right)\right)\\
&+\frac{1}{4}Q\left(\left(\sqrt{2\gamma_1}-\sqrt{2\gamma_2}\right)\sin\left(\frac{\pi}{4}+\varphi_1\right)\right)
\label{eq:serapproxnoma1}\\
P_{\mathrm{SER},2}&\approx Q\left(\sqrt{2\gamma_2}\sin\left(\frac{\pi}{4}-\varphi_2\right)\right)+Q\left(\sqrt{2\gamma_2}\sin\left(\frac{\pi}{4}+\varphi_2\right)\right) \label{eq:serapproxnoma2}
\end{align}
where $\varphi_i$, for $i\in\{1,2\}$, is given by
\begin{align}
    \varphi_i=\tan^{-1}\left(6M\sqrt{\frac{\hat{C}^{(i)}_3Md\log(N)}{N}}\right)
    \label{eqphase}
\end{align}
and $\hat{C}^{(i)}_3=C_3\log\left(M\left(12\gamma_i+\sqrt(d)\right)\right)$\footnote{A more accurate approximation can be obtained by considering the phase rotation for each user. However, the approximations in \eqref{eq:serapproxnoma1} and \eqref{eq:serapproxnoma2} are sufficiently accurate while finding more accurate approximations is beyond the scope of this paper.}.
\end{lem}

\begin{IEEEproof}
Authors of \cite{NOMAser} characterize the SER of the QPSK constellation for the uplink NOMA scenario. It is shown that the SER for the stronger user, user 1 in our case, is approximated by 
\begin{align}
    Q\left(\sqrt{\gamma_1}+\sqrt{\gamma_2}\right)+Q\left(\sqrt{\gamma_1}-\sqrt{\gamma_2}\right).
    \label{refsernoma}
\end{align}
In the proposed approach, we have the phase mismatch $\varphi_1$ [see~\eqref{eqphase} in Lemma~\ref{lem1}] between the final cluster centroids and the true centroids for user 1's signals. Similar to~\cite{some1995bit}, by incorporating this phase difference into each component of \eqref{refsernoma}, \eqref{eq:serapproxnoma1} can be easily derived. After detecting user 1's signal, we apply SIC and subtract the detected signal from the received signal. Detecting user 2's signal is then straightforward as it is a noisy QPSK signal with the interference of the other user canceled. Therefore, using Lemma \ref{lem1}, the SER of user 2 can be approximated by~\eqref{eq:serapproxnoma2}. This completes the proof.
\end{IEEEproof}

\begin{figure*}[t] 
\subfloat[$N = 500$ \label{F2a}]{%
\includegraphics[width=0.666\columnwidth]{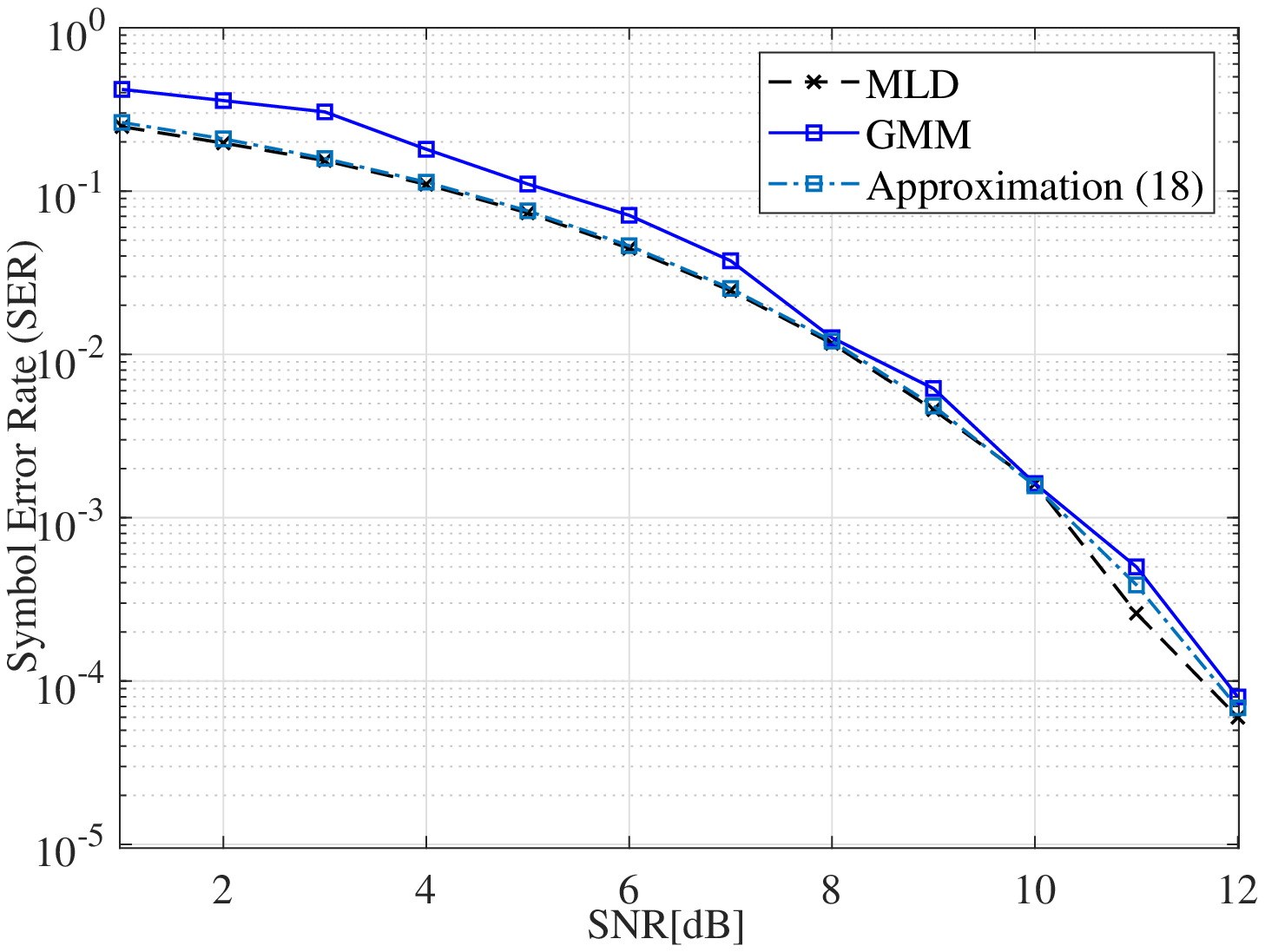}
}
\hfill
    \subfloat[$N = 200$  \label{F2b}]{%
\includegraphics[width=0.666\columnwidth]{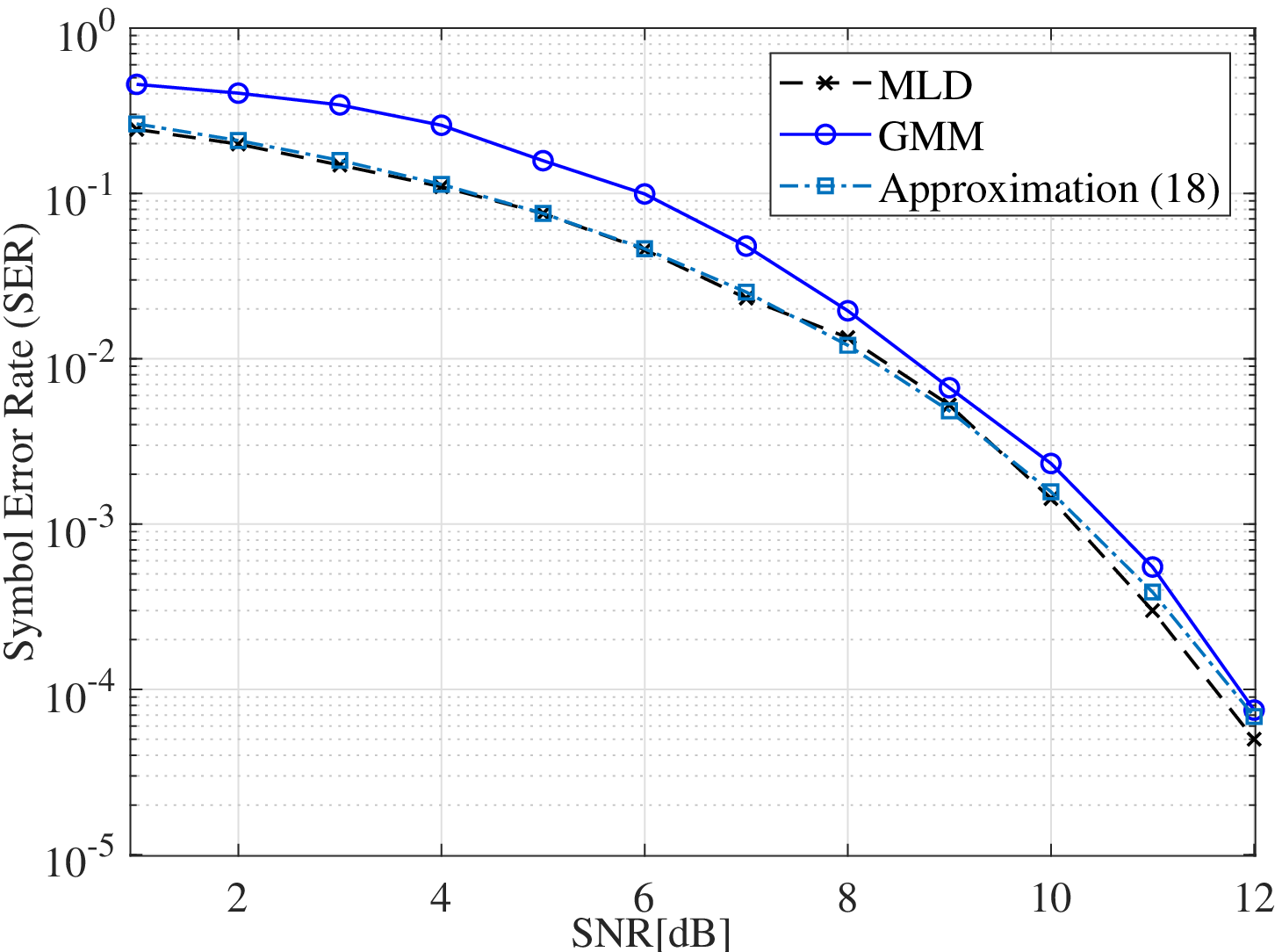}
}
\hfill
\subfloat[$N = 100$ \label{F2c}]{%
\includegraphics[width=0.666\columnwidth]{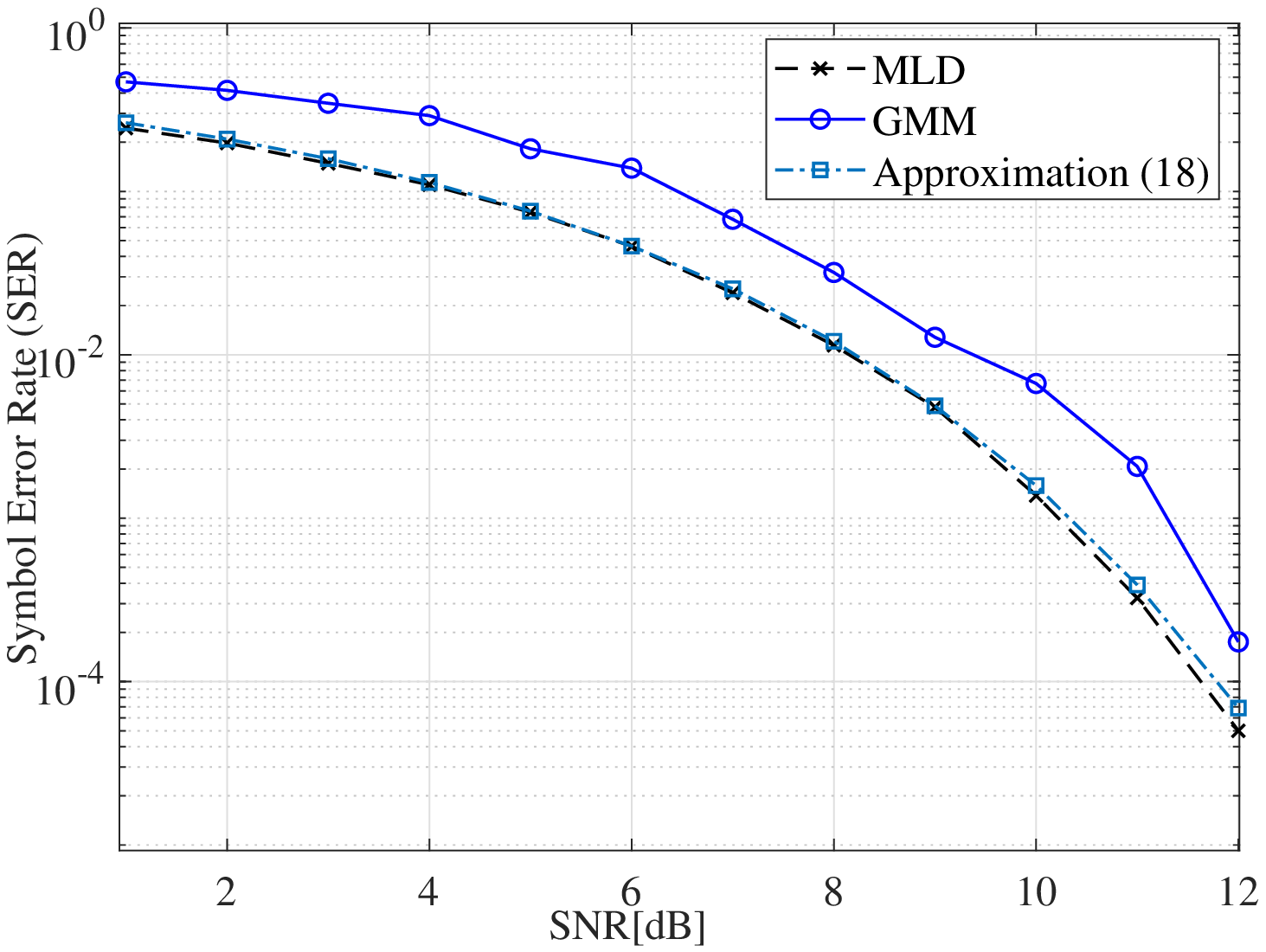}
}
\caption{\small{SER Comparison of the proposed GMM-clustering-based and optimal MLD-based approaches for point-to-point communication.}}
\label{fig:P2P-GMM}
\vspace{-1.5em}
\end{figure*}

% Note that the convergence of GMM as per Theorem~\ref{thm:1} is local since we assume that the EM algorithm is initialized in the neighborhood of the true centroids. To obtain an initial estimate, we first divide the whole data points into $M$ clusters using $k$-means clustering as mentioned in Algorithm \ref{GMM Clustering Algorithm}. We then calculate the mean of each cluster. According to our experiments, these centroids are often in the neighborhood of the true cluster centroids, and can therefore be used for initialization of the EM algorithm. When the BS does not have the knowledge of the modulation scheme, other techniques such as the method of moments~\cite{hsu2013learning} can be used for the initialization. 
Note that the convergence of GMM as per Theorem~\ref{thm:1} is local since we assume that the EM algorithm is initialized in the neighborhood of the true centroids. As the pilots are in close proximity to the true cluster centroids, they can be utilised to initialise the EM algorithm as they are at the centre of the decision boundaries.When the BS does not have the knowledge of the modulation scheme, other techniques such as the method of moments~\cite{hsu2013learning} can be used for the initialization. 

Since the EM algorithm used for GMM clustering comprises two alternating steps of expectation and maximization, to determine its computational complexity, we consider the complexity of each step. For a general case, assume we have $d$ dimensions, $M$ clusters, and $N$ samples. In the expectation step, we calculate the determinant and the inverse of the covariance matrix with the complexity of $\mathcal{O}(MNd^3)$.
%or at least $\mathcal{O}(d^{2.373})$. Therefore, the expectation costs $\mathcal{O}(MNd^3)$.
The maximizing stage entails calculating the mixture weight, mean, and covariance for each cluster with corresponding $\mathcal{O}(MN)$, $\mathcal{O}(MN)$, and $\mathcal{O}(MNd)$ complexity levels.
%The maximization step requires computing the mixture weight, mean, and covariance for each cluster with the associated complexity of $\mathcal{O}(MN)$, $\mathcal{O}(MN)$, and $\mathcal{O}(MNd)$, respectively.
Given the algorithm convergences after $t$ iterations, the computational complexity of GMM clustering is $\mathcal{O}(tMNd^3)$, which is most importantly linear in $N$. Since all computations are carried out at the BS, which generally has sufficient computing and energy resources, the mentioned computational complexity does not pose any challenge, particularly considering the substantial gains in throughput.

\section{Numerical Results}\label{Numerical Results}

In our simulations, we consider that two pilot symbols are sent for symbol-to-bit demapping, unless we specify otherwise. We also assume that the convergence threshold, $\epsilon$, for the proposed GMM-clustering-based approach is set to 1, unless specified otherwise.

\subsection{Single-User Scenario}

In Fig.~\ref{fig:P2P-GMM}, we compare the SER performance of the proposed GMM-clustering-based approach with that of the MLD-based receiver with full CSI knowledge in a single-user scenario with QPSk modulation. As seen in this figure, when the sample size $N$ is large, the proposed approach performs close to the optimal MLD-based one with full CSI, while at the high SNR regime, the SERs of the two approaches are almost the same. For small $N$, there is a small difference in the performance of the two approaches, which can mainly be attributed to limited observations making the cluster boundaries of GMM sub-optimal. In addition, Fig. \ref{fig:P2P-GMM} shows that the approximation given by~\eqref{eq:serapprox} can predict the SER of the proposed approach well.

\subsection{Two-User NOMA}

\begin{figure*}[t] 
\subfloat[$N=500$ \label{fig:NOMA-2USer-500symbol-9db}]{%
\includegraphics[width=0.666\columnwidth]{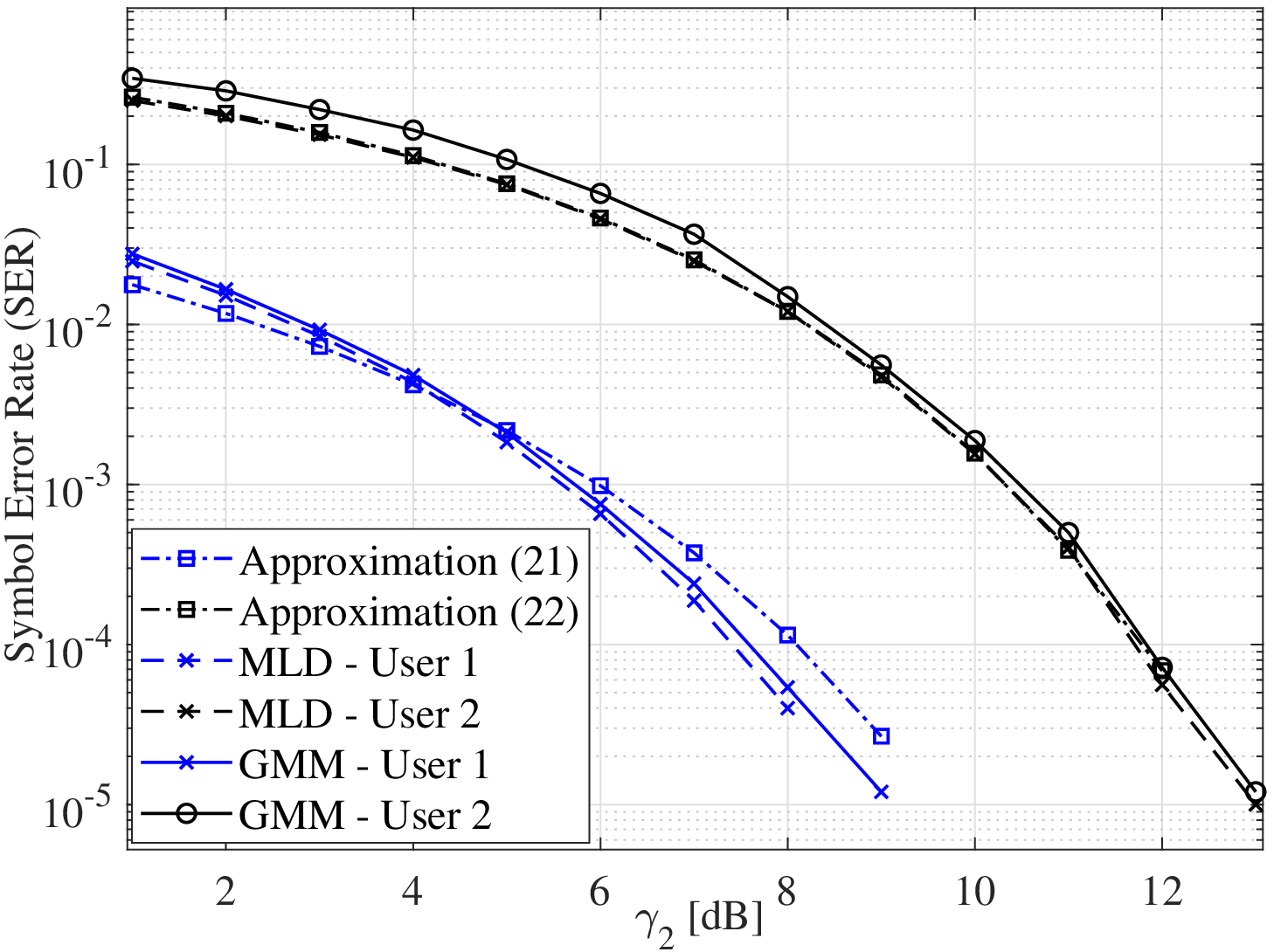}
}
\hfill
    \subfloat[$N=200$  \label{fig:NOMA-2USer-200symbol-9db}]{%
\includegraphics[width=0.666\columnwidth]{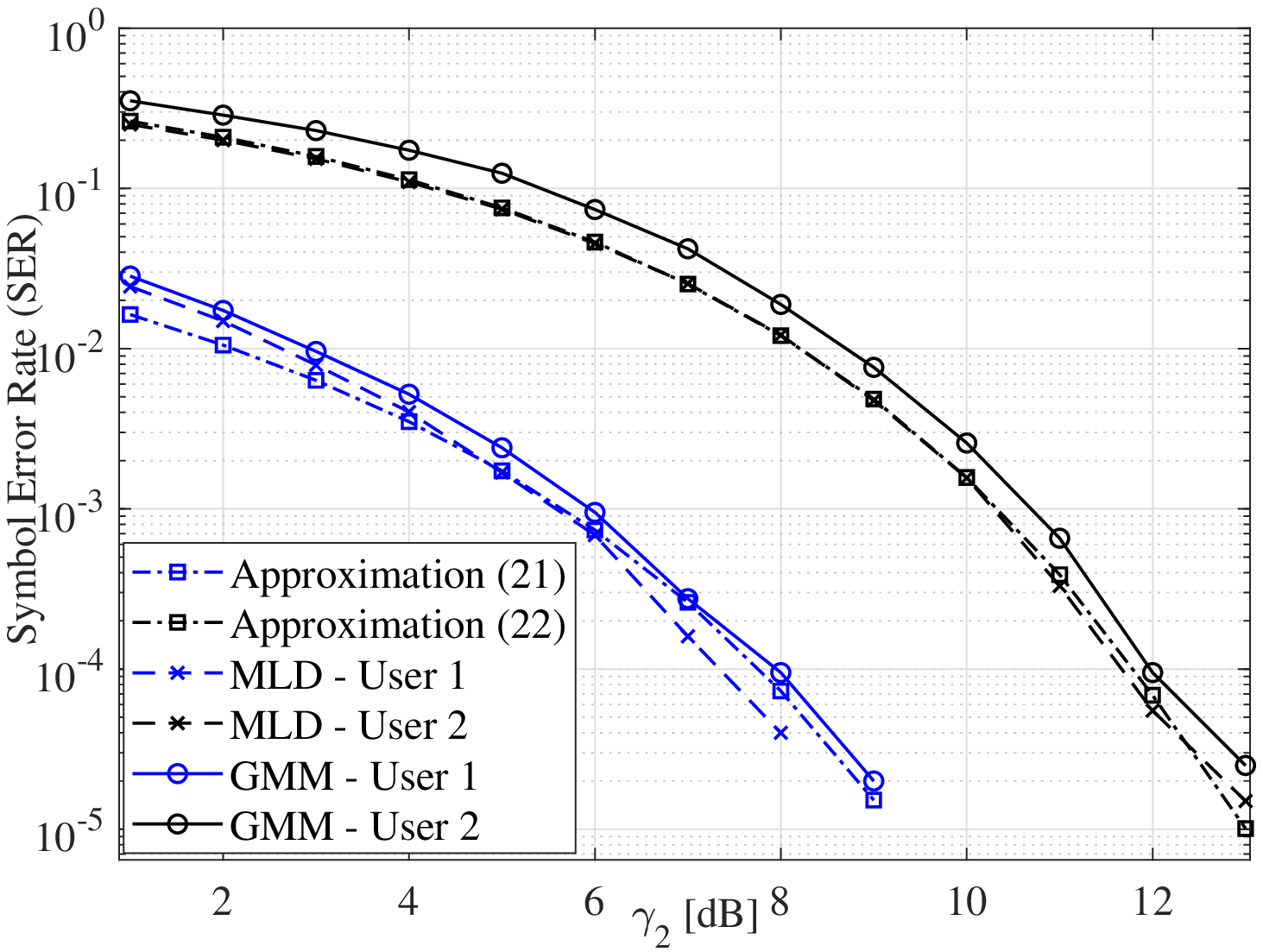}
}
\hfill
\subfloat[$N=100$ \label{fig:NOMA-2USer-100symbol-9db}]{%
\includegraphics[width=0.666\columnwidth]{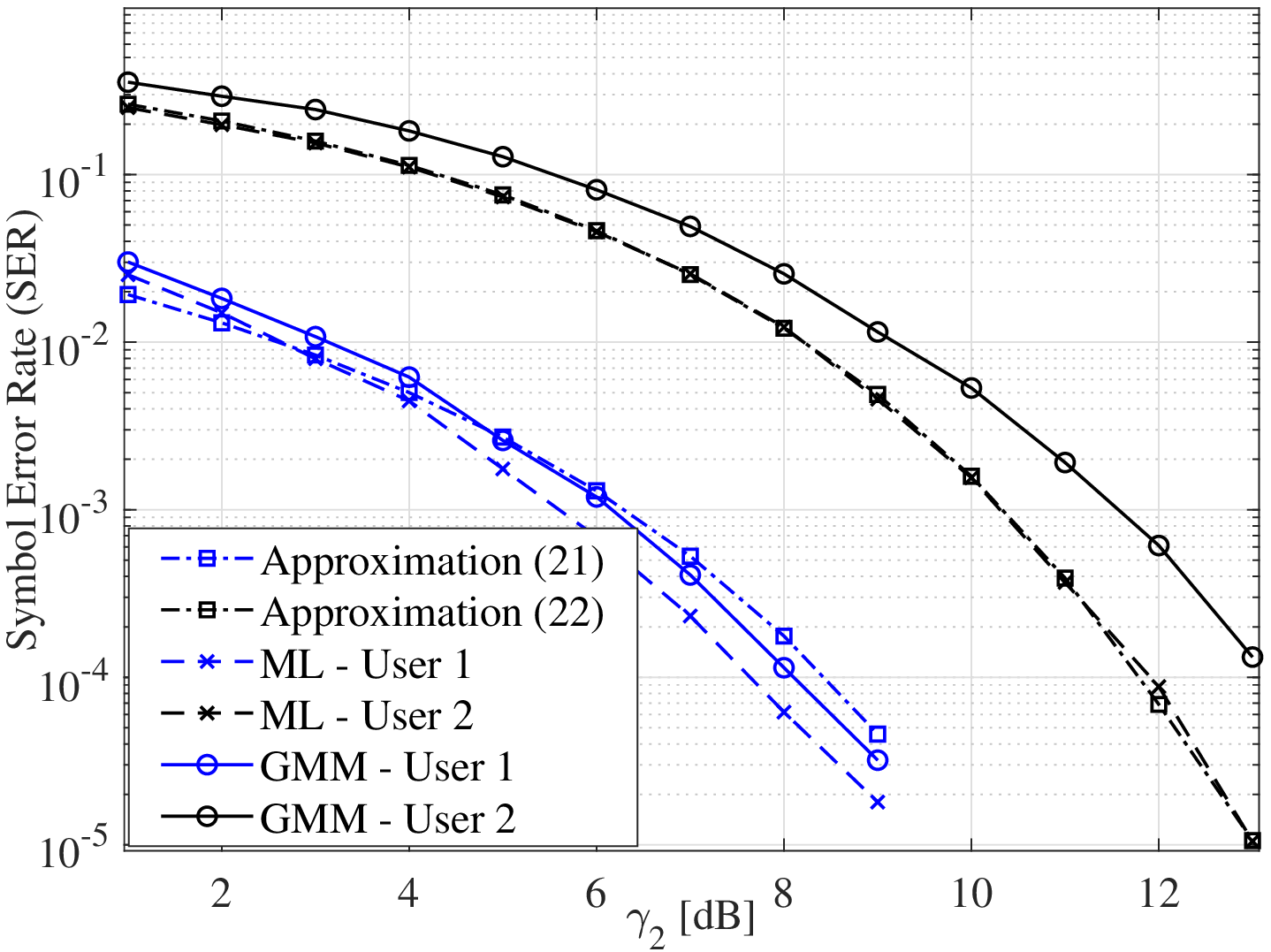}
}
\caption{\small{SER comparison of the proposed GMM-clustering-based and optimal MLD-based approaches for two-user NOMA when $\gamma_1-\gamma_2=9$dB.}}
\vspace{-1.5em}
\label{fig:2 users NOMA 9db}
\end{figure*}

We show the performance of the proposed GMM-clustering-based approach for a two-user NOMA communication system in Fig.~\ref{fig:2 users NOMA 9db}. As seen in this figure, the proposed approach performs close to the optimal MLD-based approach with full CSI. Similar to the single-user case, as the number of transmitted symbols increases, the SER performance of the proposed approach reaches that of the MLD-based approach with full CSI. An advantage of the clustering-based approach is that each user needs to send only two pilot symbols for symbol-to-bit demapping. However, in current systems, to acquire sufficiently accurate CSI at the receiver, each user needs to send a long pilot sequence that is usually more than six symbols~\cite{dong2004optimal}. This is inefficient when the packet size (number of transmitted symbols) is small~\cite{Shirvani2019Short}. In our proposed approach, the BS does not require a perfect estimation of the channel to eliminate the influence of channel rotation. We can find the exact quadrant of each cluster by using only two pilot symbols. Fig.~\ref{fig:2 users NOMA 9db} also shows that the theoretical predictions of \eqref{eq:serapproxnoma1} and \eqref{eq:serapproxnoma2} are reasonably accurate.

\subsubsection{Impact of difference in user powers}

\begin{figure*}[t] 
\subfloat[$\gamma_1 - \gamma_2 = 7$dB \label{fig:NOMA_Constellation_7dB_difference}]{%
\includegraphics[width=0.666\columnwidth]{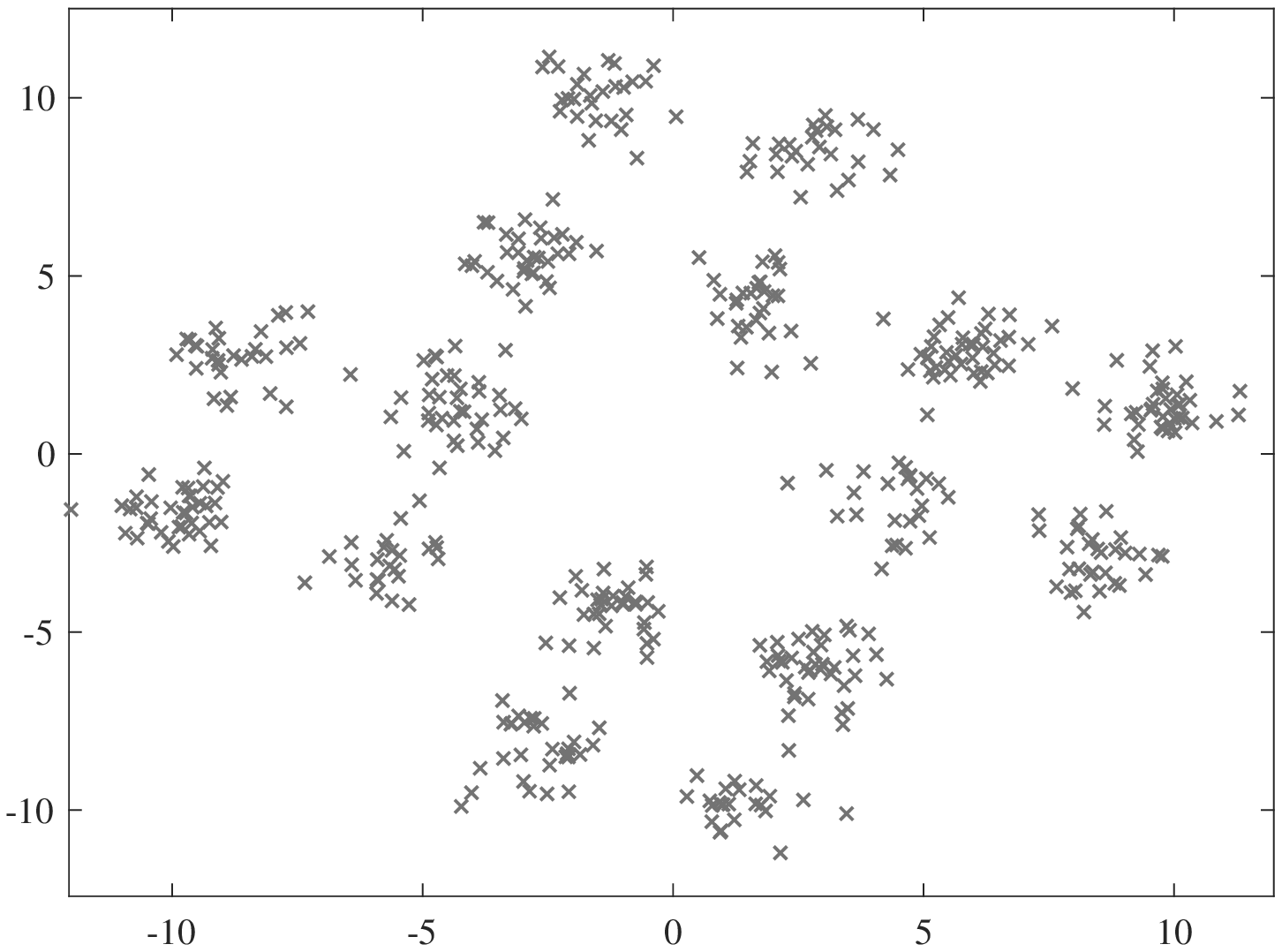}
}
\hfill
    \subfloat[$\gamma_1 - \gamma_2 = 5$dB  \label{fig:NOMA_Constellation_5dB_difference}]{%
\includegraphics[width=0.666\columnwidth]{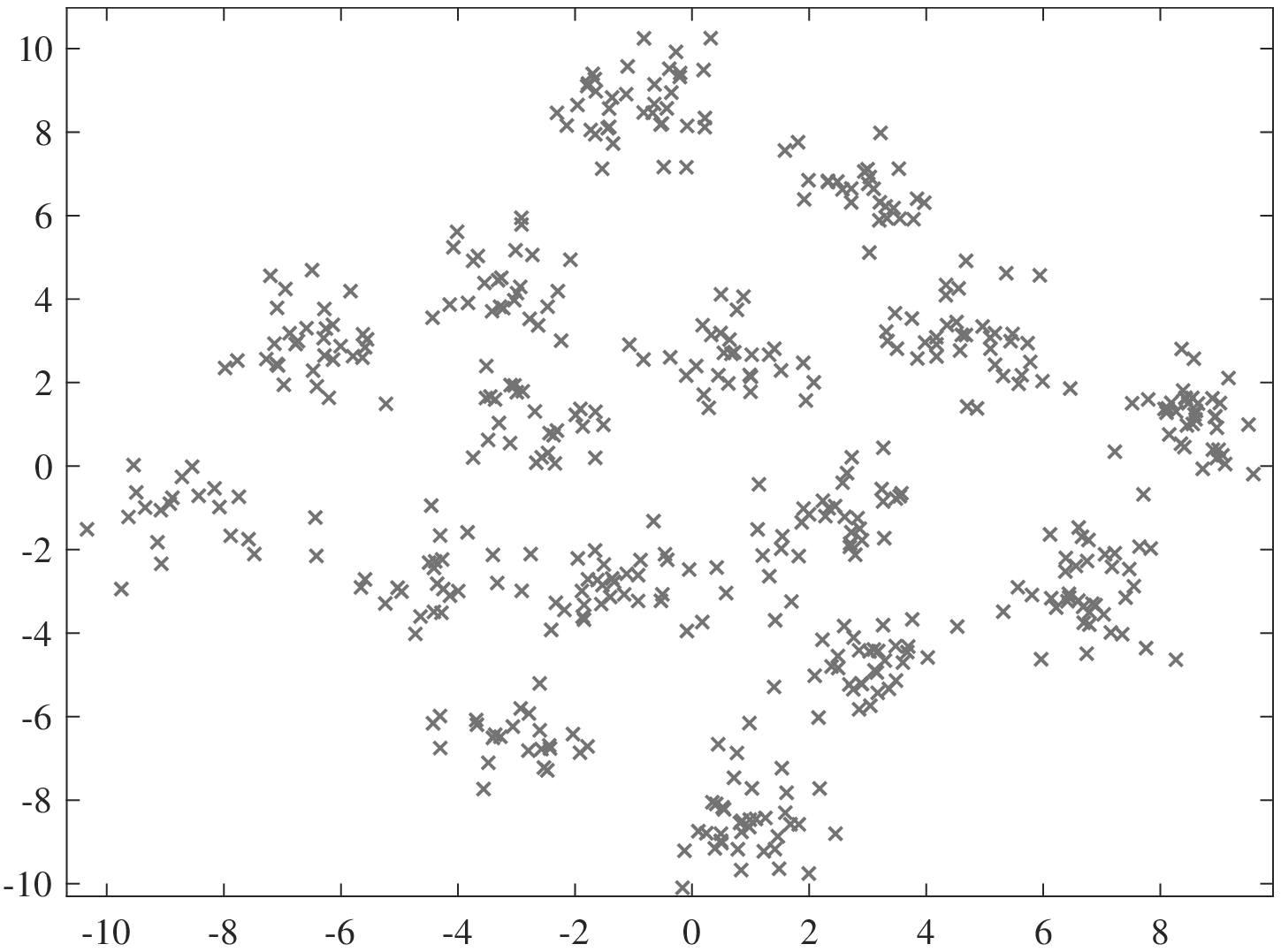}
}
\hfill
\subfloat[$\gamma_1 - \gamma_2 = 3$dB \label{fig:NOMA_Constellation_3dB_difference}]{%
\includegraphics[width=0.666\columnwidth]{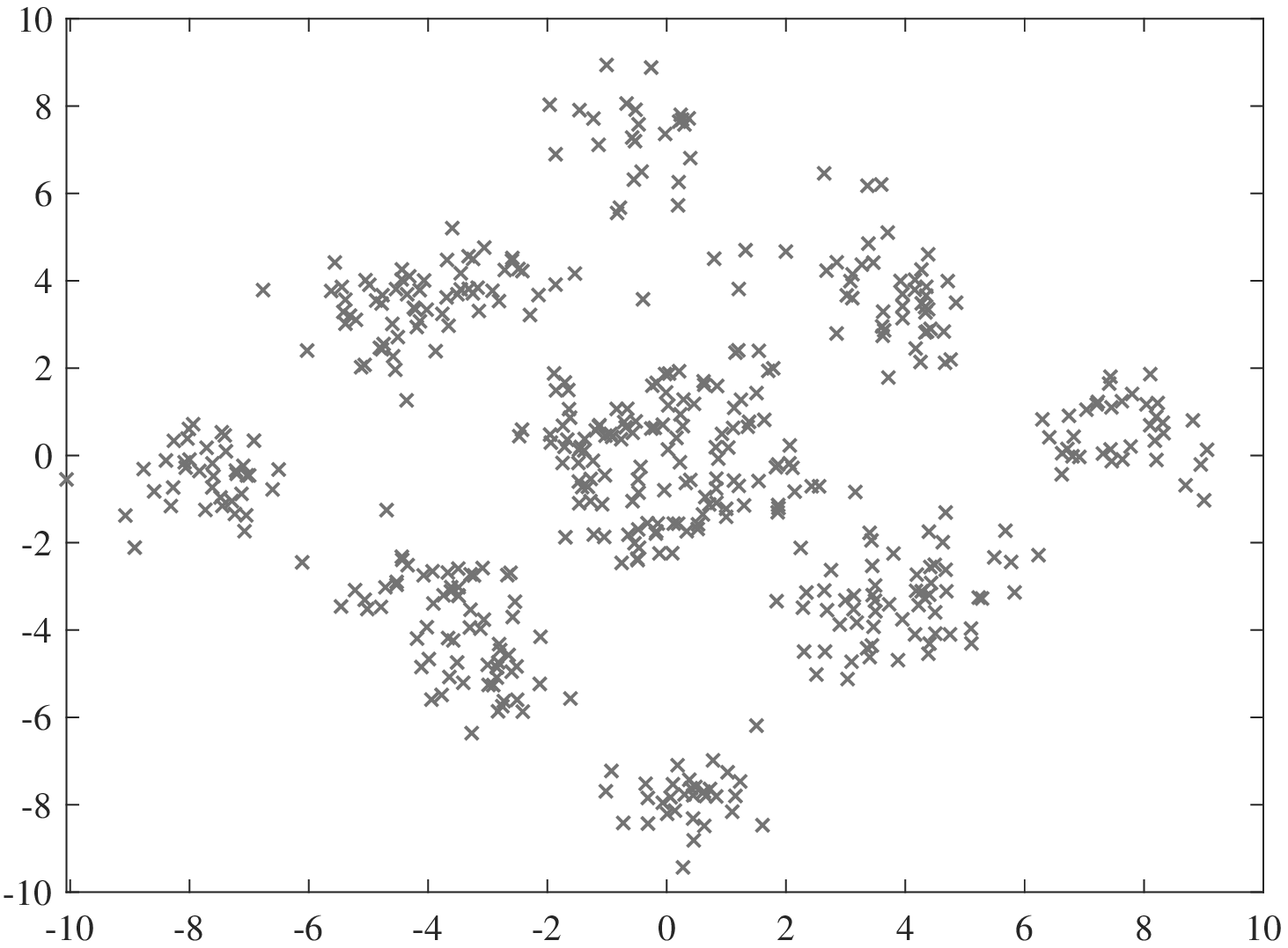}
}
\caption{\small{Constellation of received signals in a two-user NOMA scenario for three values of user power difference when $\gamma_2 =  10$dB.}}
\vspace{-1.5em}
\label{fig:NOMA_Constellation_based_on_Power}
\end{figure*}

When the received powers from the users are similar, the distance between the clusters associated with the weaker user increases. This can lead to four close clusters around the center each belonging to a different user. Distinguishing these clusters can be hard for the BS as they may be too close to each other or even overlap. Fig.~\ref{fig:NOMA_Constellation_based_on_Power} illustrates such a case. As seen in Fig.~\ref{fig:NOMA_Constellation_7dB_difference}, when the difference in the power of two users is adequately large, both users can detect their signals correctly as the received signals form distinct clusters. However, as the power difference between the users decreases (Figs.~\ref{fig:NOMA_Constellation_5dB_difference} and \ref{fig:NOMA_Constellation_3dB_difference}), the clusters associated with the weaker user(s) grow farther from each other. Consequently, four not-so-distinct clusters appear around the center that belong to different mappings.

\begin{figure*}[htp] 
\subfloat[$\gamma_1 - \gamma_2 = 7$dB \label{fig:NOMA_SER_7dB_difference}]{%
\includegraphics[width=0.666\columnwidth]{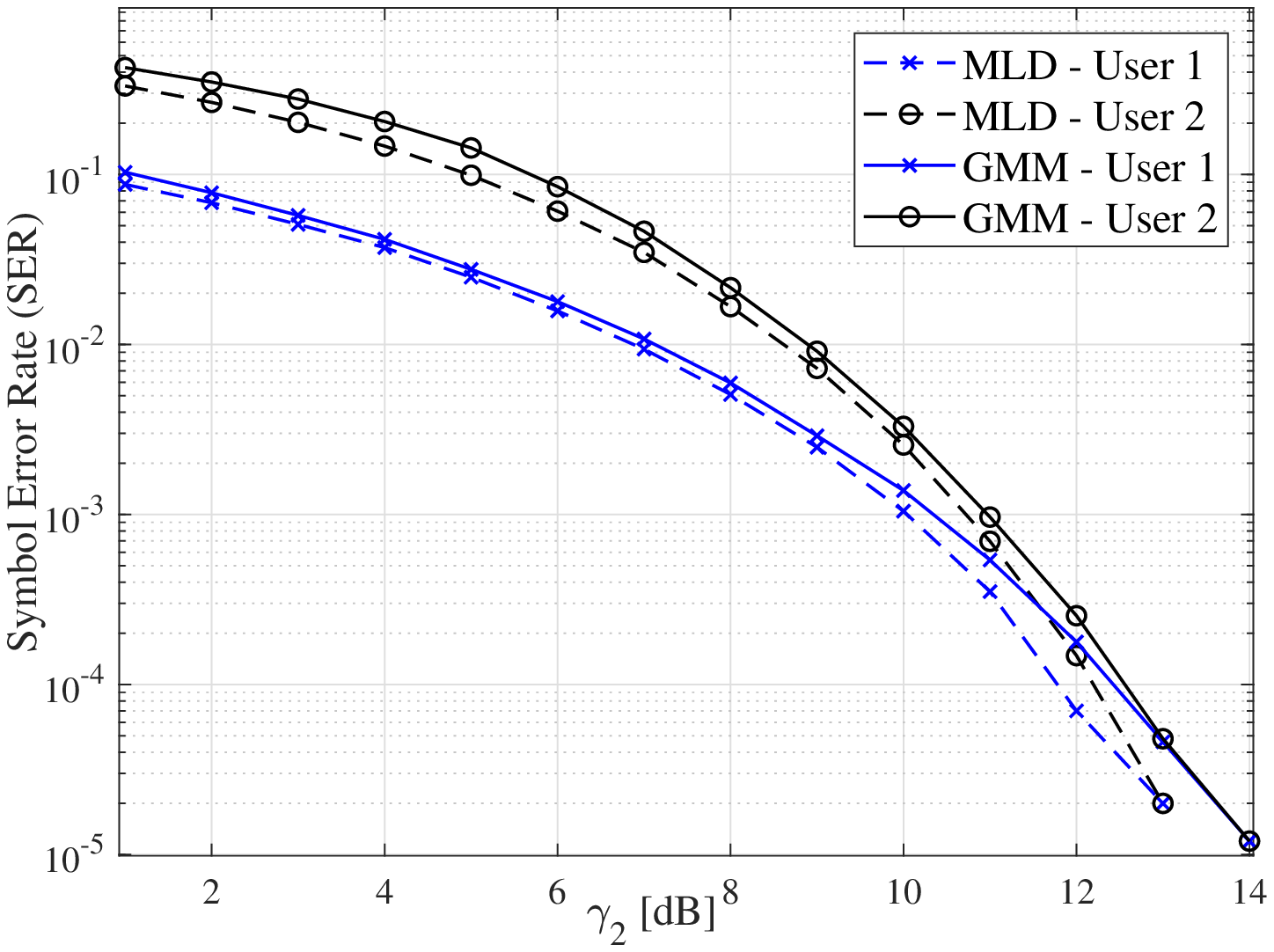}
}
\hfill
    \subfloat[$\gamma_1 - \gamma_2 = 5$dB  \label{fig:NOMA_SER_5dB_difference}]{%
\includegraphics[width=0.666\columnwidth]{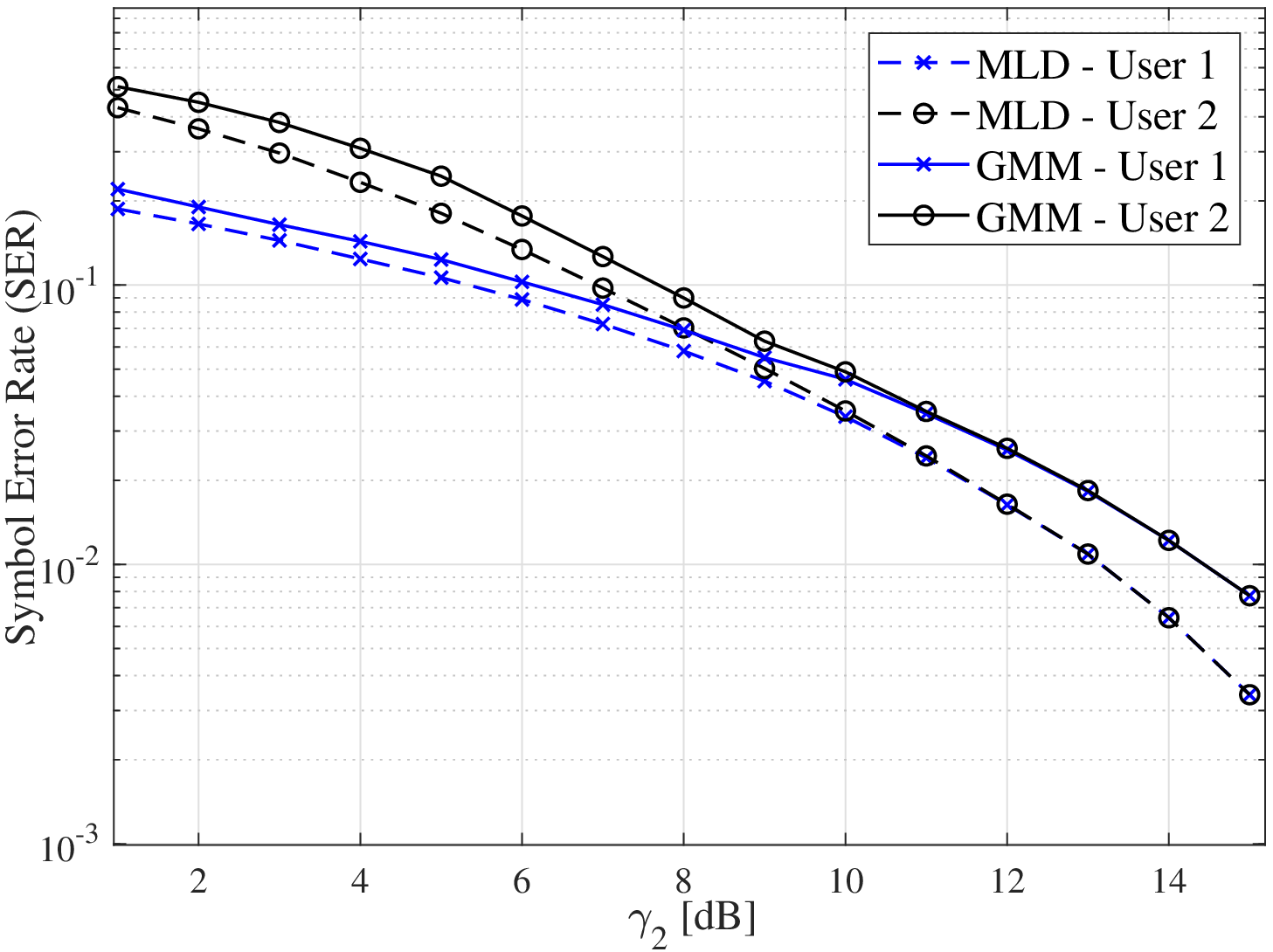}
}
\hfill
\subfloat[$\gamma_1 - \gamma_2 = 3$dB \label{fig:NOMA_SER_3dB_difference}]{%
\includegraphics[width=0.666\columnwidth]{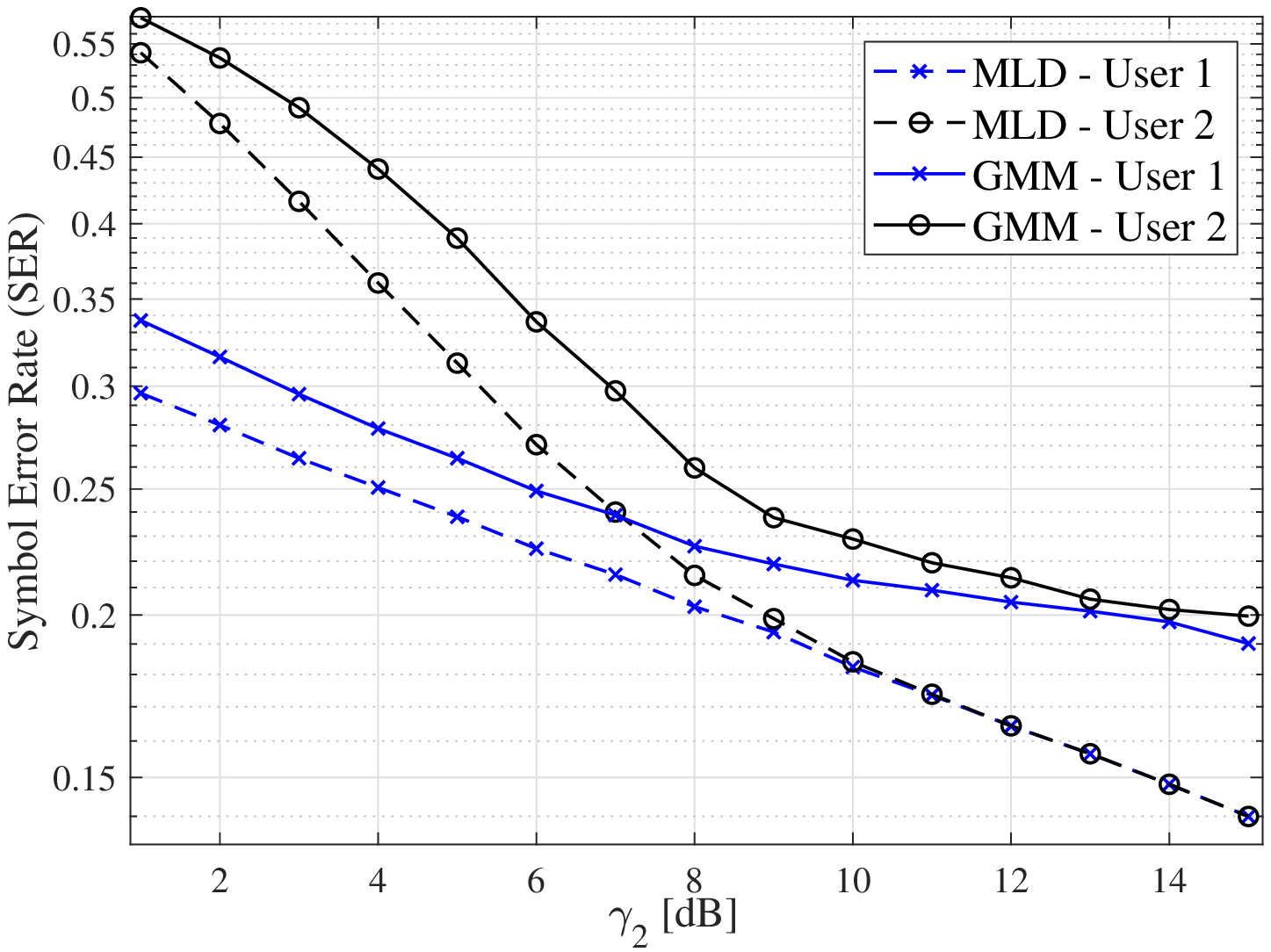}
}
\caption{\small{SER performance of the proposed GMM-clustering-based and optimal MLD-based approaches for the considered two-user NOMA scenario.}}
\vspace{-1.5em}
\label{fig:NOMA_SER_Power_Difference}
\end{figure*}

In Fig.~\ref{fig:NOMA_SER_Power_Difference}, we present the SER performance of the proposed approach in a two-user NOMA scenario for three values of user power difference. One can see from Fig.~\ref{fig:NOMA_SER_7dB_difference} that when the user power difference is sufficiently large, the BS can detect the symbols correctly even with low SNR. However, as the user power difference decreases, the SER deteriorates for both users. The SER performance of a two-user NOMA when the user power difference is relatively low is shown in Figs.~\ref{fig:NOMA_SER_5dB_difference} and Fig.~\ref{fig:NOMA_SER_3dB_difference}. We observe that the proposed algorithm can detect the signals of both users even when the users' received signal powers at the BS are close to each other. It is important to note that in power-domain NOMA, the power of the signals from different users should be substantially different. Otherwise, the BS will not be able to distinguish the signals. This is clear in Fig.~\ref{fig:NOMA_SER_3dB_difference}, which shows that, when the user power difference is 3dB, even the optimal MLD-based receiver performs poorly.

\subsubsection{Comparison with semi-blind estimation}

\begin{figure}[t]
    \centering
    \includegraphics[width=0.9\columnwidth]{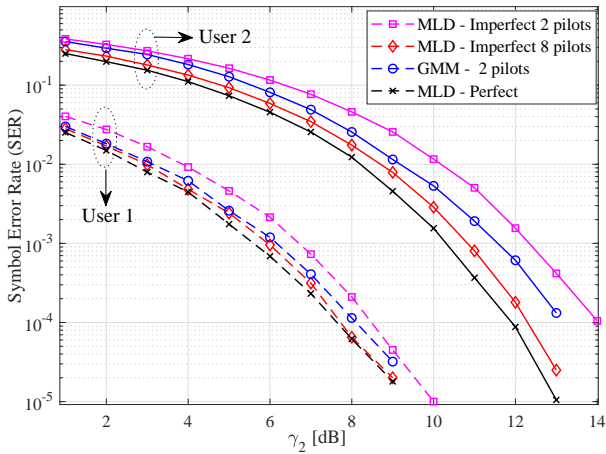}
    \caption{\small{SER performance of the proposed GMM-clustering-based approach, MLD-based approach with full CSI, and MLD-based approach with imperfect channel estimation for a two-user NOMA scenario when $\gamma_1-\gamma_2=9$dB and $N=100$.}}
    \label{fig:Imperfect ML}
%    \vspace{-1.5em}
\end{figure}

We have so far compared our proposed joint estimation and detection approach with the optimal approach based on MLD with full CSI. However, in real-world scenarios, obtaining full CSI with only a limited number of symbols is infeasible. Fig.~\ref{fig:Imperfect ML} shows the performance of the proposed algorithm when two symbols are used for demapping in comparison with the MLD-based approach with full CSI, MLD-based approach using two training symbols for channel estimation, and MLD-based approach using eight training symbols for channel estimation. As the results show, the performance of MLD with two symbols used for channel estimation is significantly inferior to the proposed algorithm using the same two pilot symbols. For the MLD-based approach to attain a performance close to that of our proposed algorithm, it requires at least eight pilot symbols to estimate the channel sufficiently accurately. This means, for a packet length of 100 symbols, the proposed algorithm offers at least six percent improvement in throughput.
%\hl{for obtaining full CSI, the channel for each user needs to be estimated separately. Did you consider this?}\textcolor{blue}{Yes} \hl{You mean by 8 symbols in total, you can obtain full CSI for both users?} \textcolor{blue}{No, I mean that 8 symbols imperfect MLE has a performance close to full CSI MLE. }

\begin{figure}[t]
    \centering
    \includegraphics[width=0.9\columnwidth]{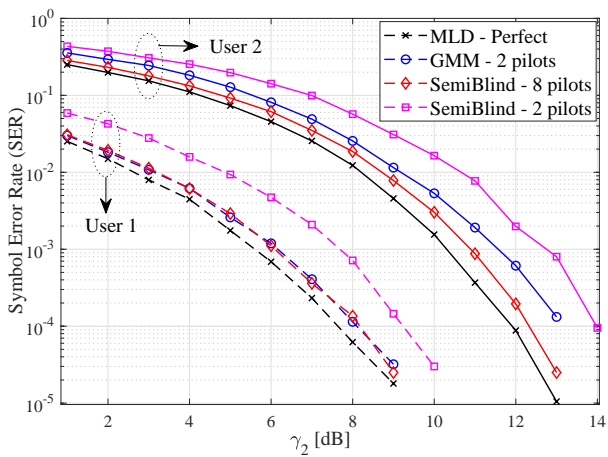}
    \caption{\small{SER performance of the proposed GMM-clustering-based approach and the one based on semi-blind channel estimation for a two-user NOMA scenario when $\gamma_1-\gamma_2=9$dB and $N=100$.}}
    \label{fig:SemiBlind}
    %\vspace{-1.5em}
\end{figure}

Fig.~\ref{fig:SemiBlind} shows the performance of our proposed algorithm in comparison with a receiver based on the semi-blind channel estimation method proposed in~\cite{murthy2006training}. We observe that, with the same number of pilot symbols, the proposed algorithm offers considerably better performance for all users. For the strongest user, the approach based on semi-blind channel estimation needs at least eight pilot symbols to perform on a par with the proposed algorithm. This means the proposed algorithm can deliver at least six percent higher throughput. Note that the semi-blind approach using ten pilot symbols may offer better SER performance for the weaker user.

\subsubsection{Impact of the convergence threshold $\epsilon$}

\begin{figure}[t]
    \centering
    \includegraphics[width=0.9\columnwidth]{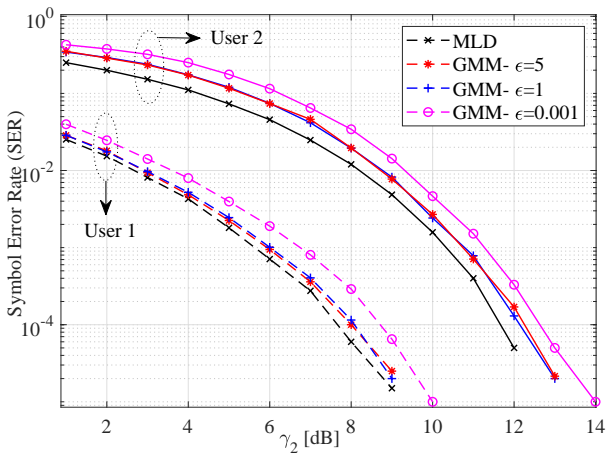}
    \caption{\small{SER performance of the proposed approach in a two-user NOMA scenario with different values of $\epsilon$ when $\gamma_1-\gamma_2=9$dB and $N=500$.}} %\hl{In MATLAB, use the latex interpreter and change eps to $\epsilon$ in the legend}}
    \label{fig:Epsilon Threshold}
%    \vspace{-1.5em}
\end{figure}

One of the parameters that should be considered when using the proposed algorithm is the convergence threshold $\epsilon$. This parameter has a direct impact on the time complexity of the proposed algorithm as well as its SER performance. As shown in Fig.~\ref{fig:Epsilon Threshold}, when the value of $\epsilon$ is small, not only the speed of the algorithm decreases, but also its performance degrades due to possible over-estimation. As the convergence threshold increases, both the SER performance and convergence speed improve. However, the threshold ought to be set carefully as increasing it excessively may result in under-estimation.% Moreover, $\epsilon$ should not be very small as the SER performance will degrade, mainly because of the overestimation. 

\subsubsection{Impact of pilot symbol number on symbol-to-bit demapping}

\begin{figure}[t]
    \centering
    \includegraphics[width=0.9\columnwidth]{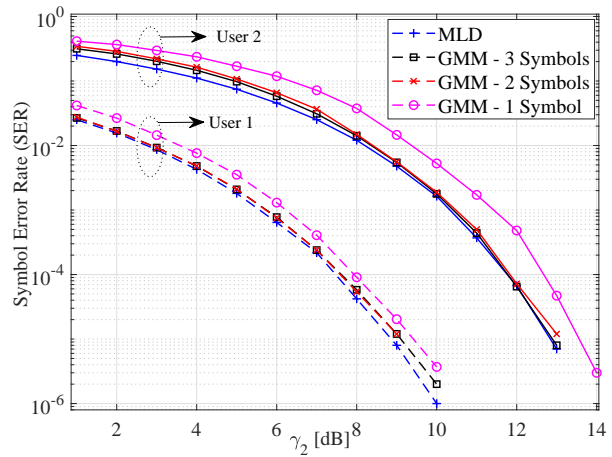}
    \caption{\small{SER performance of the proposed approach and the optimal MLD-based approach with full CSI for a two-user NOMA scenario with different numbers of pilot symbols when $\gamma_1-\gamma_2=9$dB and $N=500$.}}
    \label{fig:One Symbol Pilot}
%    \vspace{-1.5em}
\end{figure}

As we mentioned earlier, in the high SNR regime and when the received powers from the users are sufficiently different, a single pilot symbol is enough for demapping. With the QPSK modulation and using the one-symbol demapping pilot, we can identify the clusters belonging to each quadrant. Afterwards, considering the centroid of each cluster, we can calculate the phase and modulus of each channel. We demonstrate the performance of the one-symbol-based detection in Fig.~\ref{fig:One Symbol Pilot}. It is evident that, by using one symbol for demapping, the SER performance for the stronger user can approach that of the MLD-based approach with full CSI. In addition, there is less than 1dB difference between the SER performance for the weaker user and the optimal MLD-based detection.

\subsection{Higher Numbers of Users}

\begin{figure}[t]
    \centering
    \includegraphics[width=0.9\columnwidth]{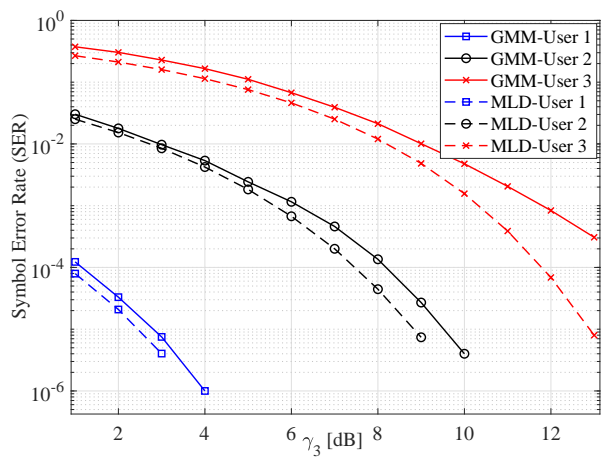}
    \caption{\small{SER performance of the proposed GMM-clustering-based approach and the MLD-based approach with full CSI for a three-user NOMA scenario when $\gamma_1-\gamma_2=9$dB, $\gamma_2-\gamma_3=9$dB, $N=500$, and $\epsilon=5$.}}
    \label{fig:3 User NOMA}
    \vspace{-0.5em}
\end{figure}

To evaluate the performance of the proposed algorithm when the number of users increases, we consider a three-user NOMA scenario. Fig.~\ref{fig:3 User NOMA} shows the SER performance of the proposed algorithm for three-user NOMA when the number of symbols is $N=500$ and three pilot symbols are used for symbol-to-bit demapping. It is clear that the proposed algorithm detects the signals with good accuracy. We note that in power-domain NOMA, usually two or three users are paired. This is because the SIC at the receiver cannot successfully detect user signals, if their powers do not differ sufficiently~\cite{8788603}. Maintaining large power differences among a large number of paired signals is impractical.
%

%\begin{algorithm}[t]
%        \KwIn{Received data at BS, number of Gaussian distributions M}
%       \KwOut{SER of GMM.}
%           %
%            Implement MMSE and utilize the only pilot symbol available to assign clusters to mapping
%           
%            \textbf{Initialize} total rotation using the MMSE estimation of the pilot
%            
%            Apply GMM clustering for user 1 and calculate the average distance of centers and their rotation
%            
%            Apply SIC and Run GMM for the second user and calculate the distance of centers from the origin
%            
%            Based on the received pilot, the rotation of the second user can be estimated
%            
%            Use the rotation value to assign clusters of the second user to mapping
%        
%
%        \caption{Applying clustering of NOMA using One symbol Pilot}
%        \label{One Symbol Clustering Algorithm}
%\end{algorithm}
%

\begin{figure}[t]
    \centering
    \includegraphics[width=0.9\columnwidth]{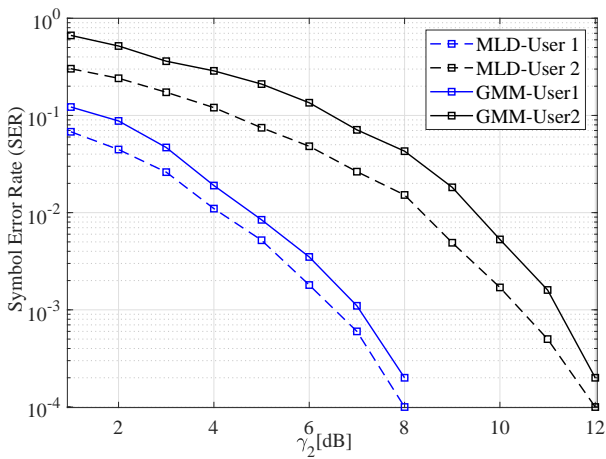}
    \caption{\small{SER performance of the proposed algorithm in a two-user NOMA scenario when user 1 and user 2 employ 16-QAM and QPSK modulation schemes, respectively, $\gamma_1 - \gamma_2 = 15$dB, and   $N=500$}.}
    \label{fig:different-modulation}
    %\vspace{-1em}
\end{figure}

\subsection{Higher-order Modulations}

To demonstrate that the proposed algorithm can be used with any modulation scheme, we consider a two-user NOMA scenario where user 1 utilizes the 16-QAM modulation scheme and user 2 utilizes the QPSK modulation scheme. As before, we benchmark the SER performance of the proposed algorithm, which uses two pilot symbols, against that of the optimal MLD-based approach with full CSI. Fig.~\ref{fig:different-modulation} shows that the proposed algorithm with only two pilot symbols performs nearly as well as the MLD-based approach with full CSI. For the MLD-based approach, the BS needs to obtain the CSI for all users using long pilot sequences. However, the proposed algorithm only needs two pilot symbols, which are used to determine the symbol-to-bit demapping.  

 %\hl{any comment whether this complexity is negligible noting to the gain in the throughput?}

\section{Grant-free Transmission: A Practical Scenario}\label{Grant-free}

Thus far, we have assumed that the number of users is fixed and the receiver knows this number. This information is used to determine the number of clusters in our proposed algorithm. In this section, we show that the proposed algorithm can be implemented in a real-world scenario where the receiver does not have any prior knowledge except for the modulation scheme used by the transmitters and their power levels.

We slightly modify the proposed GMM-clustering-based algorithm (Algorithm 1) to cope with not knowing the number of users. The modified algorithm is summarized in Algorithm~\ref{Grant Free GMM Clustering Algorithm}. In particular, we first find the average signal power $\mathbb{E}[||\mathbf{y}||^2]$. If it is larger than the noise power (line 2), we perform GMM clustering to assign the data into $M$ 
%\hl{you have been using both m and M to denote the constelltion size, please be consistent.} 
clusters and detect the user signals. After applying SIC, we reevaluate the signal power and continue the above process until the signal power falls below the noise power\footnote{There are different approaches for discovering the number of clusters such as the elbow method~\cite{bholowalia2014ebk}, X-means clustering~\cite{pelleg2000x}, and those based on certain information criteria~\cite{gupta2010detecting}.Two main information-criterion-based approaches use the Bayesian information criterion (BIC)~\cite{neath2012bayesian} and the Akaike information criterion (AIC)~\cite{bozdogan1987model}. Both of these approaches are based on the penalized likelihood estimation method. AIC is known to be prone to overfitting while BIC to underfitting. }.

\begin{algorithm}[t]
        \KwIn{Received signal $\mathbf{y}$, noise power $N_0$, modulation order $M$, Signal Constellation $\mathcal{S}$, and convergence threshold $\epsilon$}
        \KwOut{Detected user signals}

            Set $\omega_j = \frac{1}{M}, ~~j=1,\cdots,M$

         \While{ $P_y$ > $N_0$ }
           {     
                \textbf{Initialize} $\hat{\bm{\mu}}^{(0)}$ and $\hat{\mathbf{\Sigma}}^{(0)}$ by dividing the received signals on the reference coordinate system into $M$ equal sections and select one point in each section as the initial mean and set the initial covariance to one.
                
                Calculate $\hat{\gamma}_{i,j}^{(0)}$ according to (\ref{eq:5}).
                
                Calculate the log-likelihood function according to (\ref{eq:6}).
                
                Set $t=1$.
               
            \While {$ l^{(t)} - l^{(t-1)} \geq \epsilon $}
                {
                Update $\hat{\bm{\mu}}^{(t)}$ and $\hat{\mathbf{\Sigma}}^{(t)}$ using (\ref{eq:8}) and (\ref{eq:9}).
                
                Update $\hat{\gamma}_{i,j}^{(t)}$ according to (\ref{eq:5}).
                
                Update the log-likelihood function according to (\ref{eq:6}).
                }
            \textbf{Return} optimal $\hat{\bm{\mu}}$ and $\hat{\mathbf{\Sigma}}$.
            
            Calculate the phase of each cluster centroid as $\phi_i = \tan^{-1}\left(\frac{\mathrm{Im}(\hat{\mathbf{\mu}_i})}{\mathrm{Re}(\hat{\mathbf{\mu}_i})} \right)$.
            
            Calculate the average phase as $\phi = \frac{\sum_{i=1}^{M} {\phi_i} - M \pi}{M} $ and channel amplitude $|\hat{h}_u|=\frac{1}{M}\sum_{i=1}^M \mathrm{abs}(\mu_i)$.
            
            Update the decision boundaries using $\phi$ and utilize the pilot symbols to map each cluster into the mapping bits.
            
            Demodulate the signal to symbols: $\hat{\mathbf{x}}_u=\mathrm{demod}(\mathbf{y})$.
            
            %Estimate the channel phase rotation using the pilot symbols and \eqref{eq:MMSE}
            
            Re-modulate the user signal, multiply by the estimated channel, and subtract from the received signal:
            $\mathbf{y}\leftarrow\mathbf{y}-|\hat{h}_u|e^{j\phi} \hat{\mathbf{x}}_u$ and update $P_y$.
        }
        \caption{GMM-clustering-based joint channel estimation and signal detection for an unknown number of users.}
        \label{Grant Free GMM Clustering Algorithm}
        %\vspace{-1em}
\end{algorithm}
%
%\hl{Please read this carefully and make sure this matches with what you have implemented.}
We consider a grant-free scenario where the number of active users in not known to the BS. Each user randomly selects a power level from a pool of possible values $\mathbb{P}_p=\{P_1, P_2, \cdots, P_5\}$. We assume each user performs power control such that its average received power at the BS equals the selected power level. The received signal at the BS at time instant $i$ is then given by
\begin{equation}
    y=\sum_{i=1}^{K}\sqrt{P_i}g_i x_{i}+w,
\end{equation}
where $p_i\in \mathbb{P}_p$, $g_i$ represents small-scale fading, $w\sim\mathcal{CN}(0,1)$, and $\mathbb{E}[||x_{i}||^2]=1$. The BS does not have any prior knowledge of $K$, $g_i$, or $p_i$, but knows $\mathbb{P}_p$, $N_0$, and the modulation scheme utilized by the users.   

\begin{figure}[t]
    \centering
    \includegraphics[width=0.9\columnwidth]{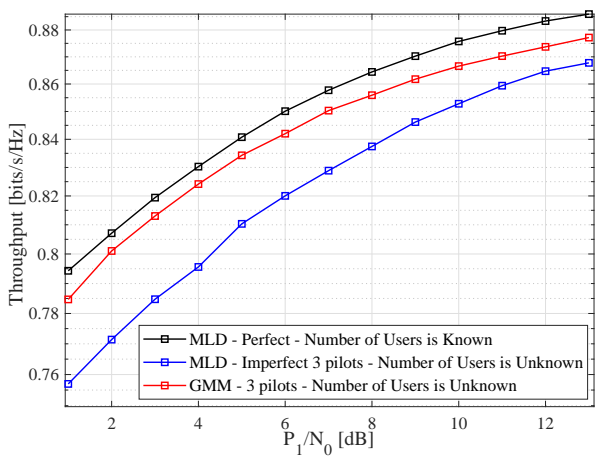}
    \caption{\small{Throughput performance of the proposed GMM-clustering-based approach versus the MLD-based one with full CSI for Grant-free transmission with Rayleigh fading when $N=500$, $\epsilon=5$, and $N_0 = 1$.}}
    \label{fig:Grant Free}
    % \vspace{-0.5em}
\end{figure}

We consider there exists small-scale fading modelled by a Rayleigh distribution, that is, $g_i\sim\mathcal{CN}(0,1)$. We also consider five power levels calculated as $P_i=P_1+(i-1)\times 5$dB for $i=1,\cdots,5$. The number of active users in each time frame (with the duration of one packet of length $N$ symbols) is uniformly distributed between one and three. At each time-slot, a random number of users start transmitting their signals. Fig.~\ref{fig:Grant Free} shows the throughput performance of the grant-free transmission with a SIC receiver based on Algorithm~\ref{Grant Free GMM Clustering Algorithm} and its comparison with the MLD-based receiver with full CSI. 
%As mentioned above, we consider five power levels and each user randomly selects one of the power levels for communication with the BS. 
At the receiver, we apply Algorithm~\ref{Grant Free GMM Clustering Algorithm} and calculate the average throughput based on the number of users. The average throughput is defined as the ratio of the number of symbols correctly  detected at the receiver and the total number of transmitted symbols. As seen in Fig.~\ref{fig:Grant Free}, the performance of the proposed algorithm with only three pilot symbols is close to that of the MLD-based approach with full CSI that has the knowledge of number of users. For a fair comparison, we also simulated the case where MLD is imperfect, BS  lacks information regarding the number of users, and the number of pilots is the same as our proposed GMM-based model. As can be seen, our proposed algorithm has significantly superior performance for the identical simulation setup.
%
%\hl{would it be possible to simulate the ML scenario with inaccurate CSI?}
%
% \begin{figure}[t]
%     \centering
%     \includegraphics[width=0.9\columnwidth]{Grant_Free.eps}
%     \caption{\small{SER performance of the proposed GMM-clustering-based approach versus the MLD-based one with full CSI for Grant-free transmission with Rayleigh fading when $N=1000$ and $\epsilon=5$.}}
%     \label{fig:Grant Free}
% %    \vspace{-1.5em}
% \end{figure}
%
%
%
%
\section{Conclusion} \label{Conclusion}
In this paper, we proposed to employ a semi-supervised machine learning algorithm, i.e., GMM clustering, to cluster the signals at the receiver for joint channel estimation and signal detection in grant-free NOMA. We applied an SIC strategy to detect the signals of each user. We compared the performance of our proposed GMM-clustering-based algorithm with that of the optimal detection method based on MLD, which requires full CSI at the receiver. We showed that using a single pilot symbol for each user, the proposed algorithm can reach the performance of the MLD-based approach with full CSI. We took one step further and used a single pilot symbol to estimate the channel of all users. The resulting performance was still close to that of the MLD with full CSI. To make a fairer comparison, we also compared the performance of our proposed algorithm with that of the MLD with imperfect channel estimation relying on a finite number of symbols. We observed that, to attain a similar performance, the MLD-based approach needs at least eight pilot symbols for channel estimation to perform as well as our proposed algorithm using only two pilot symbols. Furthermore, to reduce the computational burden at the BS, we proposed a theoretical model to calculate the probability of error based on the maximum error of the EM algorithm utilized for GMM clustering. Our simulation results showed that the proposed model predicts the SER of the proposed algorithm well. The simulation results also demonstrated that, when the number of transmitted symbols is moderate or large, the SER performance of the proposed algorithm is on a par with that of the optimal MLD. Finally, since the accuracy of the GMM clustering depends on the sample size, we showed that there exists a tradeoff between the accuracy of the proposed algorithm and the communication block length.
%\vspace{-1em}

\appendices

\section{Proof of Theorem 1}

\label{proofthm1}
The proof is based on the findings of~\cite{vershynin2018high,zhao2020statistical,yan2017convergence}.
Without loss of generality, we focus on the update rule for one of the centers.
We start by writing the update rule for the mean as
\begin{equation} 
    \bm{\mu}_1^{+} - \bm{\mu}_1^{*} =\frac{\mathbb{E}[ \gamma_1(X,\bm{\mu}) (X-\bm{\mu}_1^{*}) ] }{\mathbb{E}[ \gamma_1(X,\bm{\mu}) ]}.
\end{equation} 
Since the vector of the true centers $\bm{\mu}^{*}$ is fixed, we have 
\begin{equation} 
    \mathbb{E}[ \gamma_1(X,\bm{\mu}_1^{*}) (X-\bm{\mu}_1^{*}) ] = 0.
\end{equation} 
Hence, we can write
\begin{equation} \label{u}
    \bm{\mu}_1^{+} - \bm{\mu}_1^{*} =\frac{\mathbb{E}\left[ \left(\gamma_1(X,\bm{\mu})-\gamma_1(X,\bm{\mu}^{*}) \right) (X-\bm{\mu}_1^{*}) \right] }{\mathbb{E}[ \gamma_1(X,\bm{\mu}) ]}.
\end{equation} 
We find an upper bound on the norm of the expectation in the numerator. Therefore, we define $$\bm{\mu}^{t}:= \bm{\mu}^{*} + t(\bm{\mu}-\bm{\mu}^{*})$$ $$g_X(t) := \gamma_1(X,\bm{\mu}^{t}).$$ Subsequently, we have
\begin{align} 
    \gamma_1(X,\bm{\mu})-\gamma_1(X,\bm{\mu}^{*}) & = \int_0^1{g_X^{\prime}(t)} dt \nonumber \\
    & = \int_0^1 {	\bigtriangledown_{\bm{\mu}} \gamma_1(x,\bm{\mu}_1^{t})^T (\bm{\mu}_1^{t} - \bm{\mu}_1^{*}) dt}.
\end{align} 
Computing the integration and applying the expectation, the upper bound can be written as
\begin{align} \label{eq:upperbound_expectation}
    & \|\mathbb{E}\left[ \left(\gamma_1(X,\bm{\mu}) -\gamma_1(X,\bm{\mu}^{*}) \right)  (X-\bm{\mu}_1^{*}) \right] \|_2 \nonumber \\
    & \leq V_1 \| \bm{\mu}_1 -\bm{\mu}_1^{*}  \|_2  + \sum_{i \neq 1} V_i \| \bm{\mu}_i -\bm{\mu}_i^{*}  \|_2 \\
    & \leq M \left(  \max_i V_i  \right) \left( \max_i  \| \bm{\mu}_i -\bm{\mu}_i^{*}  \|_2\right) 
\end{align} 
where
\begin{equation} 
    \resizebox{250pt}{!}{$ V_1 = \sup_{t \in [0,1] } \| \mathbb{E}\left[ \gamma_1(X;\bm{\mu}^{t}) (1-\gamma_1(X;\bm{\mu}^{t})) (X-\bm{\mu}_1^{*})  (X-\bm{\mu}_1^{t})^T\right] \|_\textit{op} $}
\end{equation} 
\begin{equation} 
    V_i = \sup_{t \in [0,1] } \| \mathbb{E}\left[ \gamma_1(X;\bm{\mu}^{t}) \gamma_i(X;\bm{\mu}^{t}) (X-\bm{\mu}_1^{*})  (X-\bm{\mu}_i^{t})^T\right] \|_\textit{op}.
\end{equation} 

Considering $Z$ as the label of $X$, one can write
\begin{equation} 
    \mathbb{E} \left[ \gamma_1(X;\bm{\mu}^{*})  \right] = \mathbb{E} \left[ \mathbb{P}_{\bm{\mu}^{*}} (Z=1 \mid X)  \right] = \omega_1 > \kappa.
\end{equation}
When $\bm{\mu}$ is in the vicinity of $\bm{\mu}^{*}$, we have $\mathbb{E} \left[ \gamma_1(X;\bm{\mu})  \right] \approx \mathbb{E} \left[ \gamma_1(X;\bm{\mu}^{*})  \right] > \kappa$. According to \cite[Lemma 5.2]{zhao2020statistical}, we know that, as long as $R_{\min} \geq 30 \min \{M,d \}^{0.5}$ and 
\begin{equation} 
     \resizebox{250pt}{!}{$ a \geq \frac{1}{2} R_{\min} - \min\{M,d\}^{0.5} \max\{4\sqrt{2} [\log(\frac{R_{\min}}{4})]_+^{0.5}, 8\sqrt{3}, 8\log(\frac{4}{\kappa})  \} $} ,
\end{equation} 
for any $\bm{\mu}_i \in \mathcal{B}(\bm{\mu}_i^{*},a) \hspace{1mm} i \in [M]$, we have $\mathbb{E} \left[ \gamma_i(X;\bm{\mu})\right] \geq \frac{3}{4} \kappa \hspace{2mm} i \in [M]$. \\
Using the above result and the upper bound \eqref{eq:upperbound_expectation}, we have
\begin{align} 
    \| \bm{\mu}_1^{+} - \bm{\mu}_1^{*} \|_2 & =\frac{ \| \mathbb{E}\left[ \left(\gamma_1(X,\bm{\mu})-\gamma_1(X,\bm{\mu}^{*}) \right) (X-\bm{\mu}_1^{*}) \right] \|_2}{\mathbb{E}[ \gamma_1(X,\bm{\mu}) ]} \nonumber \\
    & \leq \frac{4M}{3\kappa}  \left(  \max_i V_i  \right) \left( \max_i  \| \bm{\mu}_i -\bm{\mu}_i^{*}  \|_2\right).
\end{align} 

Defining the event $\mathcal{E}_{1,i}$ as 
\begin{align} 
& \sup_{\bm{\mu} \in \mathcal{U}} \| \frac{1}{n} \sum_{j=1}^{n} \gamma_i(X_j;\bm{\mu}) (X_j-\bm{\mu}_i^{*}) - \mathbb{E}[\gamma_i(X;\bm{\mu}) (X-\bm{\mu}_i^{*}) ]  \|_2 \nonumber 
\end{align}
\begin{align}
 \leq 1.5 R_{max} \left( \frac{\hat{C_3} M d \log n}{n} \right)^{0.5}
\end{align} 
and the event $\mathcal{E}_{2,i}$ as 

\begin{equation} 
\sup_{\bm{\mu} \in \mathcal{U}} \mid \frac{1}{n} \sum_{j=1}^{n} \gamma_i(X_j;\bm{\mu})  - \mathbb{E}[\gamma_i(X;\bm{\mu})]  \mid \leq  \left( \frac{\hat{C_2} M d \log n}{n} \right)^{0.5}
\end{equation} 
where
$$\mathcal{U} = \prod_{i=1}^M \mathcal{B}(\bm{\mu}^{*}, R_{\max})$$
$$\hat{C_2} = C_2 \log \left( M\left( 2\sqrt{2}R_{\max} + \sqrt{d} \right)  \right)$$
$$\hat{C_3} = C_3 \log \left( M\left( 6 R_{\max}^2 + \sqrt{d} \right)  \right),$$
and ${C_2}$ and ${C_3}$ are universal constants.

In view of~\cite[Lemmas $5.3$ and $5.4$]{zhao2020statistical}, the event $\{ \cap_{i \in [M]} \mathcal{E}_{1,i} \} \cap \{ \cap_{i \in [M]} \mathcal{E}_{2,i} \}$ for all $i \in [M]$ occurs with the probability at least $1-\frac{2M}{n}$.
Due to the sample size condition \eqref{eq:sample_constraint}, we have 
\begin{equation} 
R_{\max} \left( \frac{\hat{C_3} M d \log n}{n} \right)^{0.5} \leq \frac{\kappa}{3} \underset{ i \in [M] }{\max} \hspace{1mm} \| \bm{\mu}_i^{0} - \bm{\mu}_i^{*} \|_2
\end{equation} 
\begin{equation} 
     \left( \frac{\hat{C_2} M d \log n}{n} \right)^{0.5} \leq \frac{\kappa}{12}.\label{2e}
\end{equation} 
Using the definition of the event $\mathcal{E}_{2,i}$ for all $i \in [M]$, the second inequality \eqref{2e} can be written as
\begin{equation} 
\sup_{\bm{\mu} \in \mathcal{U}} \mid \frac{1}{n} \sum_{j=1}^{n} \gamma_i(X_j;\bm{\mu})  - \mathbb{E}[\gamma_i(X;\bm{\mu})]  \mid \leq \frac{\kappa}{12}.
\end{equation}

Taking $\bm{\mu}^{(0)}$ as the initial value of the mean, we can write
\begin{align} 
    & \| \bm{\mu}_i^{(1)} - \bm{\mu}_i^{*} \|_2  = \frac{\| \frac{1}{n} \sum_{j=1}^{n} \gamma_i(X_j;\bm{\mu}^{(0)}) (X_j-\bm{\mu}_i^{*})   \|_2}{\frac{1}{n} \sum_{j=1}^{n} \gamma_i(X_j;\bm{\mu}^{(0)}) } \label{a}\\
    & \leq \frac{ \| \mathbb{E}[\gamma_i(X;\bm{\mu}^{(0)}) (X-\bm{\mu}_i^{*}) ]  \|_2 + R_{\max} \left( \frac{\hat{C_3} M d \log n}{n} \right)^{0.5}}{\mathbb{E}[\gamma_i(X;\bm{\mu}^{(0)})] - \frac{\kappa}{12}} \label{b}\\
    & \leq \frac{ \| \mathbb{E}[\gamma_i(X;\bm{\mu}^{(0)}) (X-\bm{\mu}_i^{*}) ]  \|_2 + R_{\max} \left( \frac{\hat{C_3} M d \log (n)}{n} \right)^{0.5}}{\frac{2 \kappa}{3}} \label{c}\\  
    & \leq \frac{1}{2}\underset{ i \in [M] }{\max} \hspace{1mm} \| \bm{\mu}_i^{(0)} - \bm{\mu}_i^{*} \|_2 +  \frac{3  R_{\max} \left( \hat{C_3} M d  \frac{\log n}{n} \right)^{0.5}}{2 \kappa}. \label{d}
\end{align} 
Using the above inequality and the first sample size condition, we have 
\begin{align} 
    \| \bm{\mu}_i^{(1)} - \bm{\mu}_i^{*} \|_2 & \leq \frac{1}{2}\underset{ i \in [M] }{\max} \hspace{1mm} \| \bm{\mu}_i^{(0)} - \bm{\mu}_i^{*} \|_2 +  \frac{3  R_{\max} \left( \hat{C_3} M d  \frac{\log (n)}{n} \right)^{0.5}}{2 \kappa}\\
    & \leq \underset{ i \in [M] }{\max} \hspace{1mm} \| \bm{\mu}_i^{(0)} - \bm{\mu}_i^{*} \|_2.
\end{align} 
By applying \eqref{a}-\eqref{d} over $t$ iterations, we have

\begin{align} 
    & \underset{ i \in [M] }{\max} \| \bm{\mu}_i^{(t)} - \bm{\mu}_i^{*} \|_2 \nonumber \\ & \leq \resizebox{245pt}{!}{$ \frac{1}{2^t}\underset{ i \in [M] }{\max} \hspace{1mm} \| \bm{\mu}_i^{(0)} - \bm{\mu}_i^{*} \|_2 +  \left( 1 + \frac{1}{2} + \cdots + \frac{1}{2^{t-1}} \right)\frac{3  R_{\max} \left( \hat{C_3} M d  \frac{\log (n)}{n} \right)^{0.5}}{2 \kappa} $}
\end{align} 
\begin{align}
     \leq \frac{1}{2^t}\underset{ i \in [M] }{\max} \hspace{1mm} \| \bm{\mu}_i^{(0)} - \bm{\mu}_i^{*} \|_2 + \frac{3  R_{\max} \left( \hat{C_3} M d  \frac{\log (n)}{n} \right)^{0.5}}{ \kappa}.
\end{align} 

%\end{IEEEproof}

\bibliographystyle{IEEEtran}
\bibliography{IEEEabrv,main_Revision1}

% Generated by IEEEtran.bst, version: 1.14 (2015/08/26)
\begin{thebibliography}{10}
\providecommand{\url}[1]{#1}
\csname url@samestyle\endcsname
\providecommand{\newblock}{\relax}
\providecommand{\bibinfo}[2]{#2}
\providecommand{\BIBentrySTDinterwordspacing}{\spaceskip=0pt\relax}
\providecommand{\BIBentryALTinterwordstretchfactor}{4}
\providecommand{\BIBentryALTinterwordspacing}{\spaceskip=\fontdimen2\font plus
\BIBentryALTinterwordstretchfactor\fontdimen3\font minus
  \fontdimen4\font\relax}
\providecommand{\BIBforeignlanguage}[2]{{%
\expandafter\ifx\csname l@#1\endcsname\relax
\typeout{** WARNING: IEEEtran.bst: No hyphenation pattern has been}%
\typeout{** loaded for the language `#1'. Using the pattern for}%
\typeout{** the default language instead.}%
\else
\language=\csname l@#1\endcsname
\fi
#2}}
\providecommand{\BIBdecl}{\relax}
\BIBdecl

\bibitem{series2017minimum}
ITU, ``Minimum requirements related to technical performance for {IMT-2020}
  radio interface (s),'' \emph{Report ITU-R}, pp. 2410--2420, 2017.

\bibitem{Framingham2019Growth}
M.~Framingham, ``Growth in connected {IoT} device,'' \emph{International Data
  Corporation (IDC)}, 2019.

\bibitem{6GCellular}
M.~Vaezi, A.~Azari, S.~R. Khosravirad, M.~Shirvanimoghaddam, M.~M. Azari,
  D.~Chasaki, and P.~Popovski, ``Cellular, wide-area, and non-terrestrial
  {IoT}: A survey on {5G} advances and the road toward {6G},'' \emph{IEEE
  Communications Surveys \& Tutorials}, vol.~24, no.~2, pp. 1117--1174, 2022.

\bibitem{shahab2019grantfree}
M.~B. Shahab, R.~Abbas, M.~Shirvanimoghaddam, and S.~J. Johnson, ``{Grant-free
  Non-orthogonal Multiple Access for IoT: A Survey},'' \emph{IEEE
  Communications Surveys \& Tutorials}, vol.~22, no.~3, 2020.

\bibitem{aldababsa2018tutorial}
M.~Aldababsa, M.~Toka, S.~G{\"o}k{\c{c}}eli, G.~K. Kurt, and O.~Kucur, ``A
  tutorial on non-orthogonal multiple access for {5G} and beyond,''
  \emph{wireless communications and mobile computing}, 2018.

\bibitem{abbas2020grantfree}
R.~Abbas, T.~Huang, B.~Shahab, M.~Shirvanimoghaddam, Y.~Li, and B.~Vucetic,
  ``Grant-free non-orthogonal multiple access: A key enabler for {6G-IoT},''
  2020.

\bibitem{mahmood2020six}
N.~H. Mahmood, H.~Alves, O.~A. L{\'o}pez, M.~Shehab, D.~P.~M. Osorio, and
  M.~Latva-Aho, ``Six key features of machine type communication in {6G},'' in
  \emph{2020 2nd 6G Wireless Summit (6G SUMMIT)}.\hskip 1em plus 0.5em minus
  0.4em\relax IEEE, 2020.

\bibitem{letaief2019roadmap}
K.~B. Letaief, W.~Chen, Y.~Shi, J.~Zhang, and Y.-J.~A. Zhang, ``The roadmap to
  {6G}: {AI} empowered wireless networks,'' \emph{IEEE Communications
  Magazine}, vol.~57, no.~8, pp. 84--90, 2019.

\bibitem{rodrigues2019edge}
T.~K. Rodrigues, K.~Suto, and N.~Kato, ``Edge cloud server deployment with
  transmission power control through machine learning for {6G} internet of
  things,'' \emph{IEEE Transactions on Emerging Topics in Computing}, 2019.

\bibitem{kim2021deep}
J.~Kim, H.~Ro, and H.~Park, ``Deep learning-based detector for dual mode {OFDM}
  with index modulation,'' \emph{IEEE Wireless Communications Letters}, 2021.

\bibitem{tang2019future}
F.~Tang, Y.~Kawamoto, N.~Kato, and J.~Liu, ``Future intelligent and secure
  vehicular network toward {6G}: Machine-learning approaches,''
  \emph{Proceedings of the IEEE}, vol. 108, no.~2, pp. 292--307, 2019.

\bibitem{Muhammad2020Artificial}
M.~H. Siddiqui, K.~Khurshid, I.~Rashid, and A.~Ahmed~Khan, ``Artificial
  intelligence based {6G} intelligent {IoT}: Unfolding an analytical concept
  for future hybrid communication systems,'' \emph{WCSE}, pp. 26--28, 2020.

\bibitem{zhang2006optimal}
W.~Zhang, X.-G. Xia, and P.-C. Ching, ``{Optimal training and pilot pattern
  design for {OFDM} systems in Rayleigh fading},'' \emph{IEEE Transactions on
  Broadcasting}, vol.~52, no.~4, pp. 505--514, 2006.

\bibitem{shin2007blind}
C.~Shin, R.~W. Heath, and E.~J. Powers, ``{Blind channel estimation for
  {MIMO-OFDM} systems},'' \emph{IEEE Transactions on Vehicular Technology},
  vol.~56, no.~2, pp. 670--685, 2007.

\bibitem{liu2013semi}
K.~Liu, J.~P.~C. Da~Costa, H.-C. So, and A.~L. De~Almeida, ``{Semi-blind
  receivers for joint symbol and channel estimation in space-time-frequency
  {MIMO-OFDM} systems},'' \emph{IEEE Transactions on Signal Processing},
  vol.~61, no.~21, pp. 5444--5457, 2013.

\bibitem{murthy2006training}
C.~R. Murthy, A.~K. Jagannatham, and B.~D. Rao, ``Training-based and semiblind
  channel estimation for {MIMO} systems with maximum ratio transmission,''
  \emph{IEEE Transactions on Signal Processing}, vol.~54, no.~7, pp.
  2546--2558, 2006.

\bibitem{jiang2020joint}
S.~Jiang, X.~Yuan, X.~Wang, C.~Xu, and W.~Yu, ``Joint user identification,
  channel estimation, and signal detection for grant-free {NOMA},'' \emph{IEEE
  Transactions on Wireless Communications}, vol.~19, no.~10, pp. 6960--6976,
  2020.

\bibitem{wang2016joint}
B.~Wang, L.~Dai, T.~Mir, and Z.~Wang, ``Joint user activity and data detection
  based on structured compressive sensing for {NOMA},'' \emph{IEEE
  Communications Letters}, vol.~20, no.~7, pp. 1473--1476, 2016.

\bibitem{wei2016approximate}
C.~Wei, H.~Liu, Z.~Zhang, J.~Dang, and L.~Wu, ``Approximate message
  passing-based joint user activity and data detection for {NOMA},'' \emph{IEEE
  Communications Letters}, vol.~21, no.~3, pp. 640--643, 2016.

\bibitem{chen2018sparse}
Z.~Chen, F.~Sohrabi, and W.~Yu, ``Sparse activity detection for massive
  connectivity,'' \emph{IEEE Transactions on Signal Processing}, vol.~66,
  no.~7, pp. 1890--1904, 2018.

\bibitem{liu2018massive}
L.~Liu and W.~Yu, ``Massive connectivity with massive {MIMO} - part {I}: Device
  activity detection and channel estimation,'' \emph{IEEE Transactions on
  Signal Processing}, vol.~66, no.~11, pp. 2933--2946, 2018.

\bibitem{salari2020clustering}
A.~Salari, M.~Shirvanimoghaddam, M.~B. Shahab, R.~Arablouei, and S.~Johnson,
  ``Clustering-based joint channel estimation and signal detection for
  grant-free {NOMA},'' in \emph{IEEE Globecom Workshops (GC Wkshps}.\hskip 1em
  plus 0.5em minus 0.4em\relax IEEE, 2020.

\bibitem{salari2022noma}
A.~Salari, M.~Shirvanimoghaddam, M.~B. Shahab, Y.~Li, and S.~Johnson, ``{NOMA}
  joint channel estimation and signal detection using rotational invariant
  codes and {GMM-based} clustering,'' \emph{IEEE Communications Letters}, 2022.

\bibitem{rossi2018mathematical}
R.~J. Rossi, \emph{Mathematical statistics: an introduction to likelihood based
  inference}.\hskip 1em plus 0.5em minus 0.4em\relax John Wiley \& Sons, 2018.

\bibitem{murphy2012machine}
K.~P. Murphy, \emph{Machine learning: a probabilistic perspective}.\hskip 1em
  plus 0.5em minus 0.4em\relax MIT press, 2012.

\bibitem{biship2007pattern}
C.~M. Biship, ``Pattern recognition and machine learning (information science
  and statistics),'' 2007.

\bibitem{dempster1977maximum}
A.~P. Dempster, N.~M. Laird, and D.~B. Rubin, ``Maximum likelihood from
  incomplete data via the {EM} algorithm,'' \emph{Journal of the Royal
  Statistical Society: Series B (Methodological)}, vol.~39, no.~1, 1977.

\bibitem{meng1993maximum}
X.-L. Meng and D.~B. Rubin, ``Maximum likelihood estimation via the {ECM}
  algorithm: A general framework,'' \emph{Biometrika}, vol.~80, no.~2, pp.
  267--278, 1993.

\bibitem{xu2016global}
J.~Xu, D.~J. Hsu, and A.~Maleki, ``Global analysis of expectation maximization
  for mixtures of two {Gaussians},'' in \emph{Advances in Neural Information
  Processing Systems}, 2016.

\bibitem{zhao2020statistical}
R.~Zhao, Y.~Li, Y.~Sun \emph{et~al.}, ``Statistical convergence of the {EM}
  algorithm on {Gaussian} mixture models,'' \emph{Electronic Journal of
  Statistics}, vol.~14, no.~1, pp. 632--660, 2020.

\bibitem{yan2017convergence}
B.~Yan, M.~Yin, and P.~Sarkar, ``Convergence of gradient {EM} on
  multi-component mixture of {Gaussians},'' in \emph{Advances in Neural
  Information Processing Systems}, 2017.

\bibitem{ashtiani2018nearly}
H.~Ashtiani, S.~Ben-David, N.~Harvey, C.~Liaw, A.~Mehrabian, and Y.~Plan,
  ``Nearly tight sample complexity bounds for learning mixtures of {Gaussians}
  via sample compression schemes,'' in \emph{Advances in Neural Information
  Processing Systems}, 2018.

\bibitem{kwon2020algorithm}
J.~Kwon and C.~Caramanis, ``The {EM} algorithm gives sample-optimality for
  learning mixtures of well-separated {Gaussians},'' in \emph{Conference on
  Learning Theory}.\hskip 1em plus 0.5em minus 0.4em\relax PMLR, 2020.

\bibitem{yu2020joint}
B.~Yu, Y.~Cai, and D.~Wu, ``Joint access control and resource allocation for
  short-packet-based {mMTC} in status update systems,'' \emph{IEEE Journal on
  Selected Areas in Communications}, vol.~39, no.~3, pp. 851--865, 2020.

\bibitem{lv2021energy}
S.~Lv, X.~Xu, S.~Han, X.~Tao, and P.~Zhang, ``Energy-efficient secure
  short-packet transmission in {NOMA}-assisted {mMTC} networks with relaying,''
  \emph{IEEE Transactions on Vehicular Technology}, vol.~71, no.~2, pp.
  1699--1712, 2021.

\bibitem{hasan2018time}
K.~F. Hasan, C.~Wang, Y.~Feng, and Y.-C. Tian, ``Time synchronization in
  vehicular ad-hoc networks: A survey on theory and practice,'' \emph{Vehicular
  communications}, vol.~14, pp. 39--51, 2018.

\bibitem{singh2009statistical}
R.~Singh, B.~C. Pal, and R.~A. Jabr, ``Statistical representation of
  distribution system loads using {Gaussian} mixture model,'' \emph{IEEE
  Transactions on Power Systems}, vol.~25, no.~1, pp. 29--37, 2009.

\bibitem{hastie2009elements}
T.~Hastie, R.~Tibshirani, and J.~Friedman, \emph{The elements of statistical
  learning: data mining, inference, and prediction}.\hskip 1em plus 0.5em minus
  0.4em\relax Springer Science \& Business Media, 2009.

\bibitem{ester1996density}
M.~Ester, H.-P. Kriegel, J.~Sander, X.~Xu \emph{et~al.}, ``A density-based
  algorithm for discovering clusters in large spatial databases with noise.''
  in \emph{KDD}, vol.~96, no.~34, 1996.

\bibitem{ankerst1999optics}
M.~Ankerst, M.~M. Breunig, H.-P. Kriegel, and J.~Sander, ``{OPTICS}: ordering
  points to identify the clustering structure,'' \emph{ACM Sigmod record},
  vol.~28, no.~2, pp. 49--60, 1999.

\bibitem{cheng1995mean}
Y.~Cheng, ``Mean shift, mode seeking, and clustering,'' \emph{IEEE transactions
  on pattern analysis and machine intelligence}, vol.~17, no.~8, 1995.

\bibitem{hastie1996discriminant}
T.~Hastie and R.~Tibshirani, ``Discriminant analysis by {Gaussian} mixtures,''
  \emph{Journal of the Royal Statistical Society: Series B (Methodological)},
  vol.~58, no.~1, pp. 155--176, 1996.

\bibitem{some1995bit}
Y.~Some and P.~Kam, ``Bit-error probability of {QPSK} with noisy phase
  reference,'' \emph{IEE Proceedings-Communications}, vol. 142, no.~5, pp.
  292--296, 1995.

\bibitem{NOMAser}
X.~Wang, F.~Labeau, and L.~Mei, ``Closed-form {BER} expressions of {QPSK}
  constellation for uplink non-orthogonal multiple access,'' \emph{IEEE
  Communications Letters}, vol.~21, no.~10, pp. 2242--2245, 2017.

\bibitem{hsu2013learning}
D.~Hsu and S.~M. Kakade, ``Learning mixtures of spherical {Gaussians}: moment
  methods and spectral decompositions,'' in \emph{Proceedings of the 4th
  conference on Innovations in Theoretical Computer Science}, 2013.

\bibitem{dong2004optimal}
M.~Dong, L.~Tong, and B.~M. Sadler, ``Optimal insertion of pilot symbols for
  transmissions over time-varying flat fading channels,'' \emph{IEEE
  Transactions on Signal Processing}, vol.~52, no.~5, pp. 1403--1418, 2004.

\bibitem{Shirvani2019Short}
M.~{Shirvanimoghaddam}, M.~S. {Mohammadi}, R.~{Abbas}, A.~{Minja}, C.~{Yue},
  B.~{Matuz}, G.~{Han}, Z.~{Lin}, W.~{Liu}, Y.~{Li}, S.~{Johnson}, and
  B.~{Vucetic}, ``Short block-length codes for ultra-reliable low latency
  communications,'' \emph{IEEE Communications Magazine}, vol.~57, no.~2, pp.
  130--137, 2019.

\bibitem{8788603}
J.~S. Yeom, H.~S. Jang, K.~S. Ko, and B.~C. Jung, ``{BER Performance of Uplink
  NOMA With Joint Maximum-Likelihood Detector},'' \emph{IEEE Transactions on
  Vehicular Technology}, vol.~68, no.~10, pp. 10\,295--10\,300, 2019.

\bibitem{bholowalia2014ebk}
P.~Bholowalia and A.~Kumar, ``{EBK}-means: A clustering technique based on
  elbow method and k-means in {WSN},'' \emph{International Journal of Computer
  Applications}, vol. 105, no.~9, 2014.

\bibitem{pelleg2000x}
D.~Pelleg, A.~W. Moore \emph{et~al.}, ``X-means: Extending k-means with
  efficient estimation of the number of clusters.'' in \emph{Icml}, vol.~1,
  2000.

\bibitem{gupta2010detecting}
U.~D. Gupta, V.~Menon, and U.~Babbar, ``Detecting the number of clusters during
  expectation-maximization clustering using information criterion,'' in
  \emph{2010 Second International Conference on Machine Learning and
  Computing}.\hskip 1em plus 0.5em minus 0.4em\relax IEEE, 2010.

\bibitem{neath2012bayesian}
A.~A. Neath and J.~E. Cavanaugh, ``The {Bayesian} information criterion:
  background, derivation, and applications,'' \emph{Wiley Interdisciplinary
  Reviews: Computational Statistics}, vol.~4, no.~2, pp. 199--203, 2012.

\bibitem{bozdogan1987model}
H.~Bozdogan, ``Model selection and {Akaike's} information criterion ({AIC}):
  The general theory and its analytical extensions,'' \emph{Psychometrika},
  vol.~52, no.~3, pp. 345--370, 1987.

\bibitem{vershynin2018high}
R.~Vershynin, \emph{High-dimensional probability: An introduction with
  applications in data science}.\hskip 1em plus 0.5em minus 0.4em\relax
  Cambridge university press, 2018.

\end{thebibliography}

\end{document}